\documentclass[iop]{emulateapj}

\bibstyle{ApJ}
\usepackage{epsfig}
\usepackage{amsmath}
\usepackage{natbib}
\usepackage{graphicx}

\newcommand\Av  {${\rm A}_{\rm V}$}
\newcommand\Msun  {${\rm M}_\odot$}

\newcommand\Lsun  {${\rm L}_\odot$}

\newcommand\eg    {{\it e.g.\ }}

\newcommand\kms{$\rm \, km s^{-1}$}

\begin{document}

\title{Evidence for Environmental Dependence of the Upper Stellar Initial Mass Function in Orion A}
\author{Wen-Hsin Hsu\altaffilmark{1}, Lee Hartmann\altaffilmark{1}, Lori Allen\altaffilmark{2}
Jes{\'u}s Hern{\'a}ndez\altaffilmark{3}, S. T. Megeath\altaffilmark{4}, John J. Tobin\altaffilmark{1,5} and Laura Ingleby\altaffilmark{1}}
\altaffiltext{1}{Dept. of Astronomy, University of Michigan, 500 Church St., Ann Arbor, MI 48109}
\altaffiltext{2}{National Optical Astronomy Observatory, 950 North Cherry Ave, Tucson, AZ 85719}
\altaffiltext{3}{Centro de Investigaciones de Astronom{\'i}a, Apdo. Postal 264, M{\'e}rida 5101-A, Venezuela. }
\altaffiltext{4}{Dept. of Physics and Astronomy, University of Toledo, 2801 West Bancroft Street, Toledo, Ohio 43606}
\altaffiltext{5}{National Radio Astronomy Observatory, 520 Edgemont Road Charlottesville, VA 22903}
\shorttitle{The Initial Mass Function of L1641}
\shortauthors{Hsu et al.}

\begin{abstract}
We extend our previous study of the stellar population of L1641, the lower-density star-forming region of the Orion A cloud south of the 
dense Orion Nebula Cluster (ONC), with the goal of testing whether there is a statistically significant deficiency of high-mass stars in 
low-density regions. Previously, we compared the observed ratio of low-mass stars to high-mass stars with theoretical models of the stellar 
initial mass function (IMF) to infer a deficiency of the highest-mass stars in L1641.  
We expand our population study to identify the intermediate mass (late B to G) L1641 members
in an attempt to make a more direct comparison with the mass function of the nearby ONC. 
The spectral type distribution and the K-band luminosity function of 
L1641 are similar to those of the ONC \citep{Hillenbrand97,Muench02}, but problems
of incompleteness and contamination prevent us from making a detailed 
test for differences.  We limit our analysis to statistical tests of the ratio of 
high-mass to low-mass stars, which indicate a probability of only 3\%~that the 
ONC and the southern region of L1641 were drawn from the same population,
supporting the hypothesis that the upper mass end of the IMF is dependent 
on environmental density. 

\end{abstract}


\section{Introduction}
Does the formation of high-mass stars depend systematically on the environmental density of the star-forming cloud? In a recent review of 
this important question, \citet{Bastian10} found no clear evidence for variations of the stellar initial mass function (IMF) as a function of 
initial conditions, but also concluded that further study in specific local and extragalactic environments is warranted.  Even if the IMF is 
reasonably stable on a galactic scale, as a result of averaging over a variety of environments, it would still be of interest to discern IMF 
variations in specific regions as a clue to the processes of star formation.

Studies of very young regions are necessary to make a direct connection between environmental conditions and the production of massive
stars.  The most detailed investigations of the most populated nearby young clusters - the Orion Nebula Cluster (ONC) \citep{Hillenbrand97, 
Muench02, DaRio10} and NGC 2264 \citealt{Sung04} are consistent with the the canonical IMF in general and with the Salpeter slope in the 
upper-mass end in particular \citep{Bastian10}. Producing a statistically-significant test of the high-mass IMF is much more difficult in low-
density regions ($\sim$ 1 - 10 stars $pc^{-2}$), such as Taurus or Chamaeleon, simply because there are too few low-mass stars to make a
strong test (\eg \citealt{Luhman09} for Taurus and \citealt{Luhman07} for Chamaeleon I). 
Therefore, we need a survey of a large, well-populated low-density region with spectroscopic confirmation of membership to rule out 
line-of-sight contamination. It is also important to compare the high- and low-density populations directly rather than comparing with inferred 
``universal'' IMF forms, as the results may depend on the IMF form used and the statistical significance of disagreement can be difficult to 
infer \citep{Bastian10}.

The Orion A molecular cloud is an ideal site to study star-formation in both high and low-density environments. While the ONC is one of 
the most thoroughly observed young dense cluster (the stellar surface density is $\sim$ 1000 $pc^{-2}$ for the Trapezium cluster and
 $\sim$ 200 $pc^{-2}$ for the whole ONC), the outlying regions provide us with a substantial low-density population.
L1641, the low-density star-forming region south of the ONC, has a stellar population comparable to the ONC in size ($N> 1000$;
\citealt{Hsu12}; hereafter, Paper I), but has no large scale clustering and a much lower stellar surface density of $\sim$ 10 $pc^{-2}$, 
which makes it the best low-density region to study the high-mass IMF. It also has the advantage of being at approximately the same'
distance as the ONC ($\sim$~414pc; \citealt{Menten07, Kim08}), allowing the two populations to be compared directly. 

In Paper I we presented an optical spectroscopic and photometric survey of the low-mass population in L1641 to test whether the 
previously-recognized lack of high-mass stars is statistically significant. Combining the optical data to the {\it Spitzer}/IRAC survey by 
\citet{Megeath12}, we identified and spectral-typed nearly 900 members; as we are unable to observe the heavily-extincted members, we
estimated that L1641 may contain as many as $\sim$1600 stars down to 0.1\Msun. Based on the large number of low-mass stars, we 
concluded that the lack of stars earlier than B4 is inconsistent with the canonical  \citet{Chabrier05} or \citet{Kroupa01} IMFs to 
3-4 $\sigma$ significance. 

In this paper we expand our optical photometric and spectroscopic sample to the intermediate mass (late B to G) members to improve our
estimates of the mass function. We first describe our observations and data reduction in \S~\ref{sec:data}. Then, in \S~\ref{sec:results}, we
describe the observational results and attempts to use proper motions and radial velocities to reduce foreground/background contamination.  
In addition, we estimate the age of L1641 to be about 3 Myrs (see \ref{sec:age_mass}), so late O and early B stars should still be present. 
In \S~\ref {sec:comparison}, we compare the spectral type distribution and K-band luminosity function (KLF) of L1641 to that of the ONC, 
finding that we cannot make detailed tests for differences
between the two regions given problems of incompleteness and contamination. 
In \S~\ref{sec:discussion}, we relate our results to the mass of most massive star and cluster mass relation \citep{Weidner06} and discuss
the challenges in searching for density dependence of the IMF.  Our results are then summarized in \S~\ref{sec:conclusion}. 

\section{Observations and Data Reduction}\label{sec:data}

 \begin{figure}
	\begin{center}
		\includegraphics[scale=0.6]{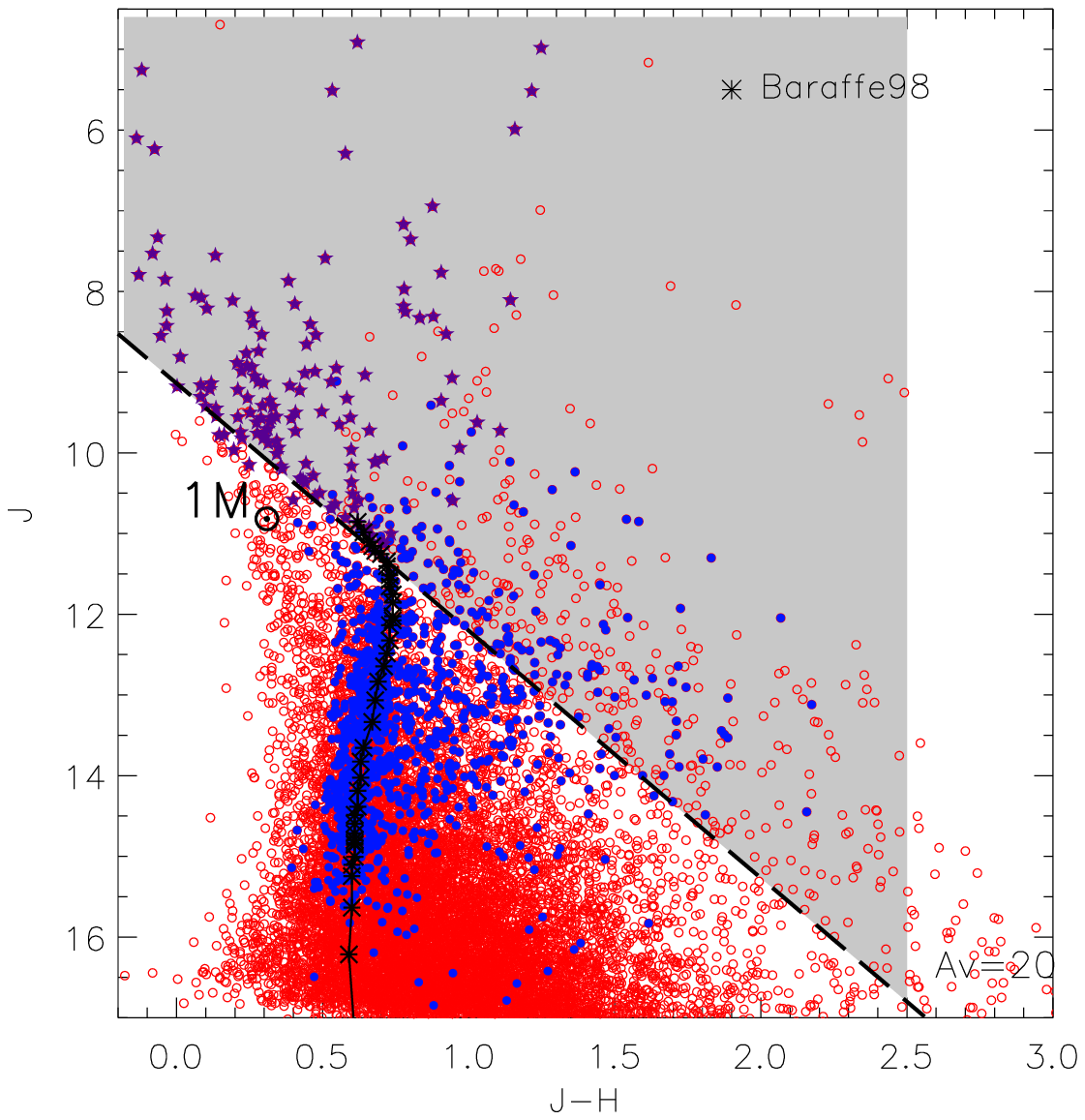}
		\includegraphics[scale=0.6]{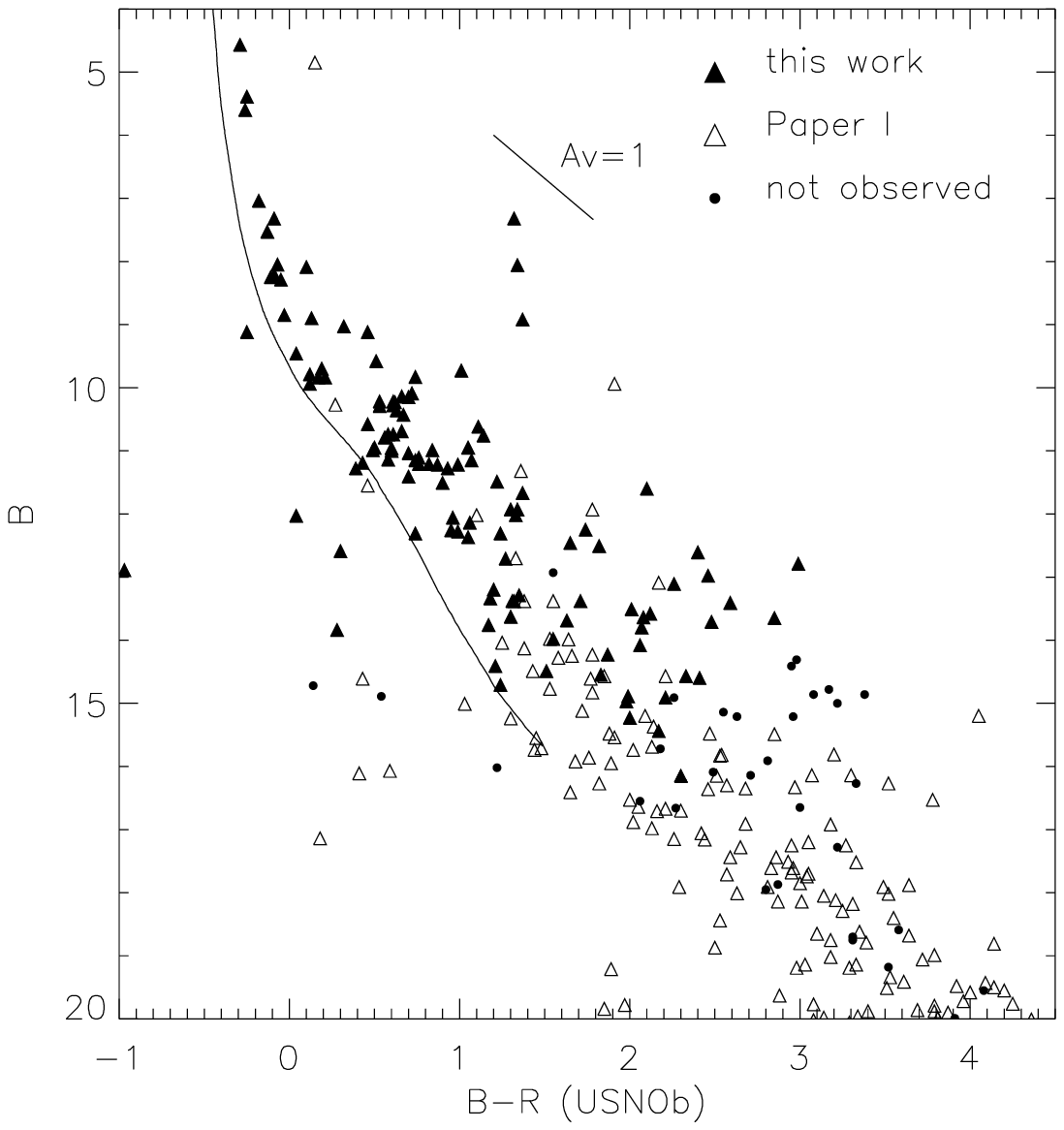}
	\end{center}
	\caption{Top: J vs. J-H color-magnitude diagram (CMD) used to select targets from the 2MASS catalog. The grey shaded region 
	indicates our initial cut for selecting targets. The blue solid circles are objects with spectral types from Paper I. The purple stars are 
	objects that we obtained spectra with OSMOS in this study.  
	Bottom: USNOb B vs. B-R CMD of targets selected in the top panel. The solid triangles are targets observed in this work, the open
	triangles are targets observed in Paper I and the the black dots are objects that were not observed.}
	\label{fig:select}
\end{figure}

\subsection{Target Selection}\label{sec:selection}
We aimed to target all the high to intermediate mass stars in L1641 with moderate extinction, regardless of whether they have disks. 
We selected our targets based on their 2MASS colors and USNOb \citep{2003AJ....125..984M} photometry because our photometric 
survey (described in PaperI) saturates at around I$\sim$12, which corresponds to late G to early K stars in L1641 in less extincted regions. 
The targets were selected using the following process: 

\begin{enumerate}
\item We plotted the J vs. J-H color-magnitude diagram (CMD) of all objects with 2MASS magnitudes in L1641 (top panel of 
Figure~\ref{fig:select}). We then selected the targets that lie above the extinction vector originating from the location of a 1\Msun~star on the 
\citet{Baraffe98} 3 Myrs isochrone. This sample included objects, if at Orion's distance, that are more massive than 1\Msun.  
\item We excluded stars that already had optical spectral types from Paper I, which left us with 446 stars. In the top panel of 
Figure~\ref{fig:select}, the blue circles show objects with spectral types from Paper I. 
\item We then compared the coordinates of our targets to the USNOb catalog. If there is not a match, then the star is too faint for our optical 
spectroscopy. We selected only objects that have B-R $<$ 2.5 and B brighter than 15 in the USNOb catalog for spectroscopic followup. 
\end{enumerate}

\begin{figure}
	\begin{center}
		\includegraphics[scale=0.6]{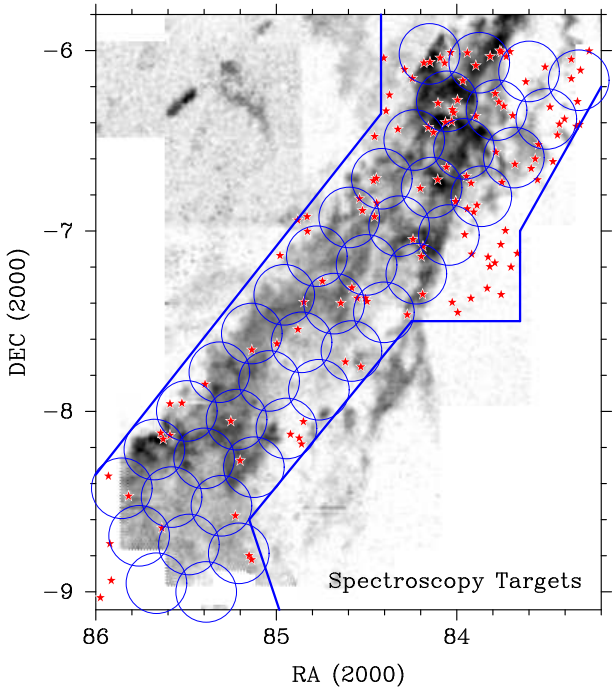}
	\end{center}
	\caption{Positions of all stars observed with OSMOS spectroscopy, overlaid on $^{13}$CO map from \citet{Bally87}. The blue
	boundary lines represent the fields covered by the {\em Spitzer}/IRAC survey of Orion \citep{Megeath12}. The circles show the fields
	of the OSMOS photometry from Paper I.}
	\label{fig:position}
\end{figure}

A total of 136 potential intermediate-mass members were observed spectroscopically (\ref{sec:optspec}). Their positions on the J vs. J-H 
CMD are shown by the stars in the top panel of Figure~\ref{fig:select}. The bottom panel of Figure~\ref{fig:select} shows the combined 
sample of stars observed in this paper with the Ohio State Multi-Object Spectrograph (OSMOS) on the MDM 2.4m \citep{Stoll10,Martini11} 
and stars from Paper I. The solid triangles are targets observed in this work, the open triangles are targets observed in Paper I. There are 
also a small number of stars that were not observed (black dots). These targets are too faint for the OSMOS spectroscopic observations. 
They were not observed with Hectospec or IMACS in Paper I because they either did not satisfy the selection criteria based on their position
on the V vs. V-I CMD, including stars that we do not have I band photometry due to saturation. 
Also shown in this figure are the extinction vector corresponding to 
\Av=1 and the ZAMS from \citet{Lejeune01} (with models from \citealt{Schaller92}).
Figure~\ref{fig:position} shows the positions of the observed stars on the sky.

\subsection{Optical Spectroscopy of Intermediate Mass Stars} \label{sec:optspec}
We obtained optical spectroscopy of selected objects (\S \ref{sec:selection}) with OSMOS. The observations were taken with center slit on 
the VPH grism, which yields a spectral range of 3100\AA~to 5950\AA~and efficiency peak at 5000\AA. The 1.4\arcsec~slit was used, giving 
a spectral resolution of R=1400, or $\sim$2 - 4\AA~for the spectral range. 

The data is first processed with the idl program proc4k written by Jason Eastman to subtract overscan and then bias subtracted, 
flat-field corrected with the MIS flat lamp to remove small-scale variation of pixel responses. Xenon comparison lamp (exposure time of
250s) is used for wavelength calibration as recommended for the center slit. Then the {\it IRAF} apall program is used to trace aperture
 and extract the spectra. 

\subsection{Optical Photometry of Intermediate Mass Stars}
We used OSMOS in photometric mode to obtain BVRI$_{\rm C}$ band photometry in the Johnson-Cousin system. The photometry
observations were all done on the night of Jan 1 - Jan 2, 2012. The conditions were photometric with seeing $\sim$ 1.2\arcsec~in the
middle of the night. 

Since the stars we are interested in are very bright stars that saturate in our previous photometric observations, we need to use an 
even shorter exposure time. (The saturation limit is I$\sim$12 for 5s exposures; see Paper I for details.) We therefore used exposure 
time of 1 second for all four bands and took another 5 second exposure for the B band image. We observed the targets once with the 
telescope in focus and then once with the telescope out of focus and the images of stars are shaped like doughnuts. This strategy 
allows us to obtain photometry of brighter stars without saturating. The FWHM of the in focus images are around 1.1 - 1.4\arcsec, 
depending on the seeing, and the out of focus images have FWHM of 2 - 2.5 \arcsec. The radius for the aperture photometry is 9 pixels, 
or 5\arcsec. 
To minimize errors due to shutter timing, we place the stars near the center of the field and use only the center 1k x 1k of the 4k x 4k 
CCD. The difference between the center and the edge of the field is about 2\% in this 1k x 1k field. The 2 x 2 binning was used, which 
gives a plate scale of 0.55\arcsec~per pixel. For calibrations, we observed Landolt fields SA92, SA95, SA98 and SA101 throughout 
the night with the full field of view of OSMOS. 

Each CCD frame was first corrected by overscan using the IDL program proc4k written by Jason Eastman. Basic reduction was performed 
following the standard procedure using {\it IRAF}. Sky flats were used in the flat-field correction. We then obtained aperture 
photometry with the {\it IRAF} $phot$ package and manually choose our target stars or the Landolt standard stars. If the 5 second B band 
exposure is saturated, we use the magnitude obtained from the 1 second exposure. The photometric errors calculated by {\it IRAF} $phot$ 
package are smaller than 0.04 mag in all cases. The rms departures of the Landolt stars from the calibration equations are $\sim$ 0.03 mag 
for all bands. We therefore expect our photometric errors to be less than 0.05 mag. 

We were unable to obtain optical photometry for stars brighter than 9th magnitude in V due to saturation. We therefore use the B \& V
photometry from the Tycho-2 catalog \citep{Tycho2}. The Tycho-2 catalog is 99\% complete to V~11.0 and the magnitude error is
0.013mag for stars brighter than 9th magnitude and 0.1mag for all other stars. The Tycho-2 photometry is transformed to the Johnson
system by using the following equation \citep{Tycho_trans}:
\begin{eqnarray}
V_J = V_T - 0.090 (B-V)_T \nonumber \\
(B-V)_J = 0.850 (B-V)_T
\end{eqnarray}

\subsection{Echelle Observations of Selected F \& G Stars}
To better distinguish foreground F \& G stars from members (see Sec~\ref{sec:member}), we observed 20 of the F \& G stars 
with MIKE, the Magellan Inamori Kyocera Echelle on the Magellan Clay telescope on Feb 4 and 5, 2012. We used both the blue side
(wavelength coverage: 3200 - 5000\AA) and the red side (wavelength coverage: 4900 - 10000\AA) simultaneously. The observations
were taken with 0.7\arcsec~slits, which gives a resolving power of 42000 in the blue side and 32500 in the red side. For the red side, we
used a milky flat taken using a hot blue star; for the blue side, we used a milky flat frame taken with the quartz lamp. We also observed ThAr
comparison lamps for wavelength correction. We reduced the data with the MIKE pipeline written by Dan Kelson and distributed as part of
the Carnegie Python Distribution. 

We then determine the radial velocities with the $rvsao.xcsao$ task in {\it IRAF}, which cross-correlates the object spectrum with a set of 
template spectra. We use synthetic spectra from Munari et al. (2005) as our input velocity templates. The templates have temperatures
from 3000 to 7000K, matching well with the temperature of the stars we observe. Each order was fitted independently, yielding a velocity
and error estimate. We then obtain the weighted mean of the velocity estimates. Typical errors for the velocity estimates are 0.5\kms~for
stars that are not rotating rapidly.

\begin{figure*}
	\begin{center}
		\includegraphics[scale=0.45]{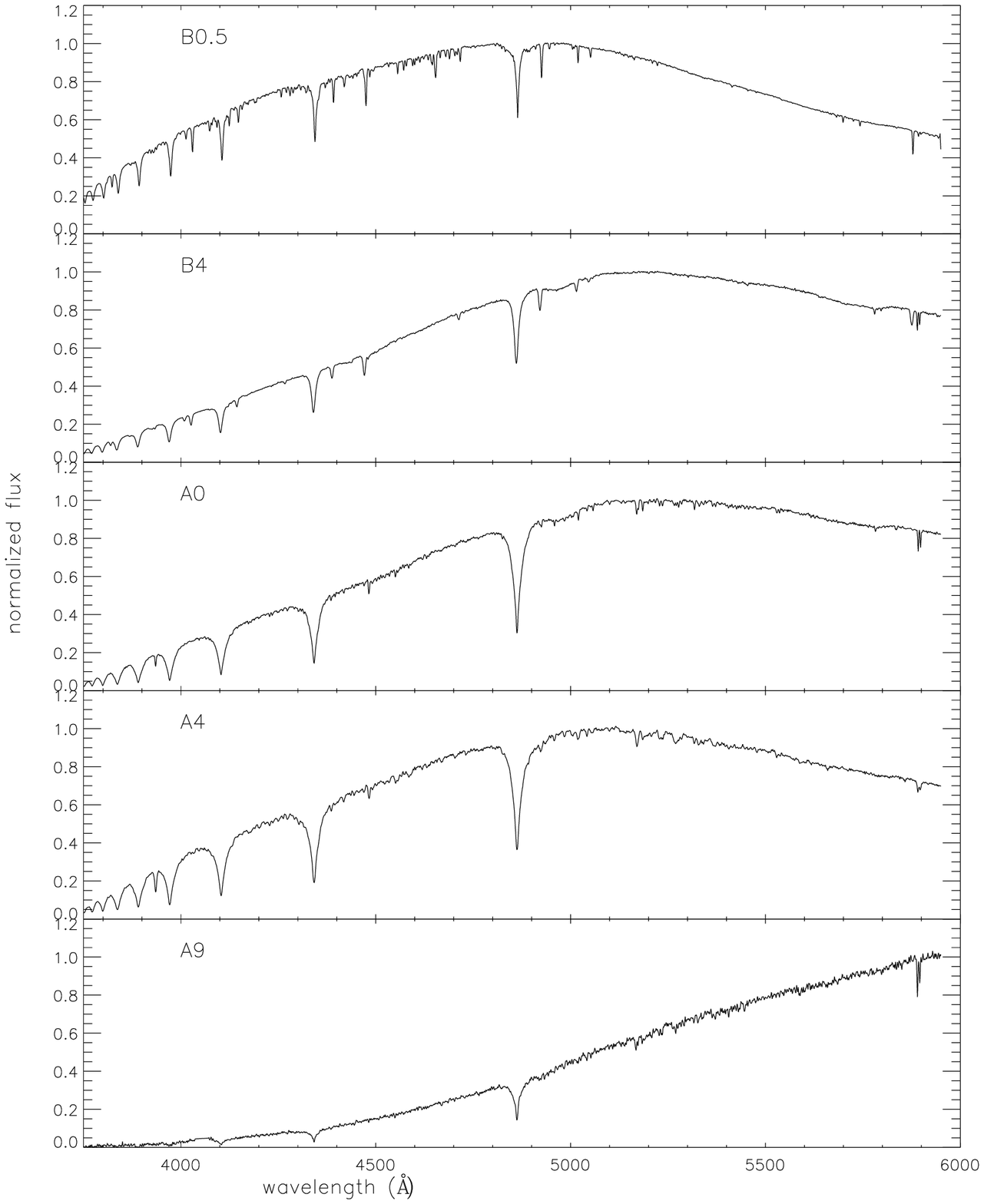}
		\includegraphics[scale=0.45]{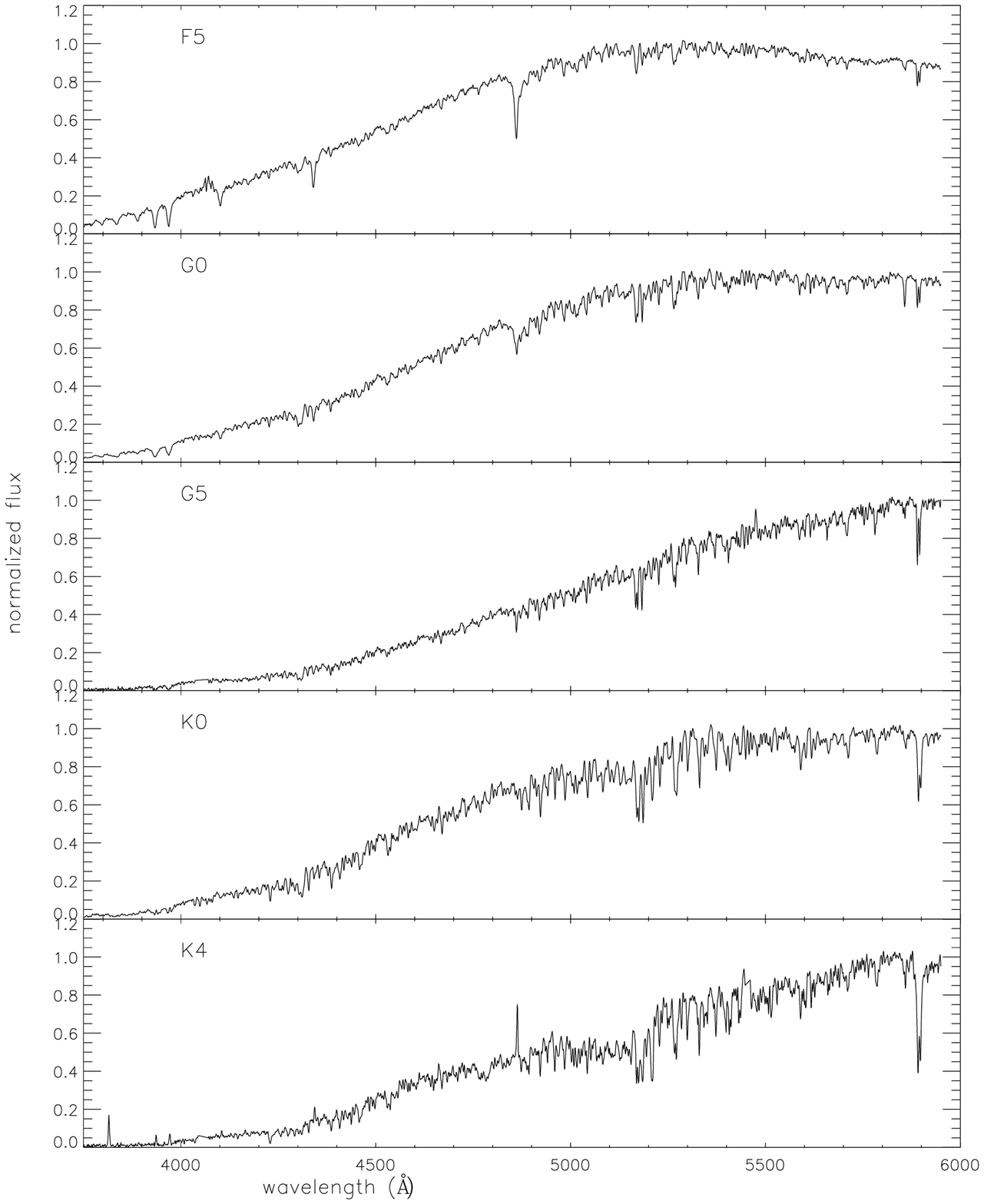}
	\end{center}
	\caption{Examples of spectra from OSMOS arranged sorted by spectral types.}
	\label{fig:spec}
\end{figure*}

\section{Results}\label{sec:results}
\begin{deluxetable*}{p{15pt}cccccccccccccc}

\scriptsize
\setlength{\tabcolsep}{0.03in} 
\tablewidth{0pt}
\tablecaption{Intermediate-mass Members of L1641}
\tablehead{\colhead{ID} & \colhead{RA} & \colhead{Dec} & \colhead{B} & \colhead{V} & \colhead{R} & \colhead{I} & \colhead{J\tablenotemark{a}} & \colhead{H\tablenotemark{a}} & \colhead{K$_S$\tablenotemark{a}} & \colhead{Spec. Type} & \colhead{IR excess\tablenotemark{b}} &\colhead{RV\tablenotemark{c}} &\colhead{EW(Li)\tablenotemark{c}} & \colhead{Note} \\
& \colhead{(J2000)} & \colhead{(J2000)} & \colhead{(mag) }& \colhead{(mag)} & \colhead{(mag)} & \colhead{(mag)} & \colhead{(mag)} & \colhead{(mag)} & \colhead{(mag)} & & & \colhead{(\kms)} &  \colhead{(\AA)} &}

\startdata
  1&  83.31498&  -6.10956&     10.65&     10.20&      9.93&      9.67&      9.32&      9.08&      8.99&F4.0$\pm$2.0& N& 17.0& 0.1& \\
  2&  83.33185&  -6.28275&     13.92&     12.72&     12.08&     11.52&     10.63&     10.09&      9.93&K1.0$\pm$3.0& N&\nodata&\nodata& \\
  3&  83.36585&  -6.04887&      9.92&      9.77&      9.71&      9.63&      9.45&      9.31&      9.29&A4.0$\pm$2.0& N&\nodata&\nodata& \\
  4&  83.43192&  -6.40788&     \nodata&     13.59&     \nodata&     12.23&     11.21&     10.53&     10.39&K3.0$\pm$3.0& N&\nodata&\nodata& \\
  5&  83.46720&  -6.61558&      9.47&      9.48&      9.49&      9.47&      9.31&      9.22&      9.18&B9.0$\pm$1.5& N&\nodata&\nodata& \\
  6&  83.56561&  -6.60124&     15.21&     14.06&     13.37&     12.46&     10.59&      9.65&      8.97&G8.0$\pm$2.0& Y&\nodata&\nodata& \\
  7&  83.61783&  -6.17349&     13.93&     12.82&     12.15&     11.47&     10.51&      9.91&      9.73&K2.0$\pm$3.0& N&\nodata&\nodata& \\
  8&  83.67800&  -6.63059&     14.07&     12.58&     11.65&     10.72&      9.49&      8.99&      8.78&F7.0$\pm$2.0& N&\nodata&\nodata& \\
  9&  83.69040&  -6.36113&     15.40&     14.06&     13.18&     12.27&     11.02&     10.35&     10.16&K4.0$\pm$2.0& N&\nodata&\nodata& \\
 10&  83.70463&  -6.00634&      8.19&      8.26&     \nodata&     \nodata&      8.55&      8.60&      8.63&B8.3$\pm$1.4& N&\nodata&\nodata& \\

\enddata
\label{tab:highmass}
 \tablenotetext{a} {\scriptsize J,H,K$_S$ photometry is from 2MASS}
 \tablenotetext{b} {\scriptsize Criteria for IR excess are defined in Gutermuth et al. (2009) and Megeath et al. (2012).}
 \tablenotetext{c} {\scriptsize From Magellan/MIKE echelle data.}
 \tablenotetext{d} {\scriptsize Herbig Ae star V380 Ori.}

\end{deluxetable*}
\begin{deluxetable*}{p{15pt}cccccccccccccc}

\scriptsize
\setlength{\tabcolsep}{0.03in} 
\tablewidth{0pt}
\tablecaption{Non-members of L1641 Observed with OSMOS}
\tablehead{\colhead{ID} & \colhead{RA} & \colhead{Dec} & \colhead{B} & \colhead{V} & \colhead{R} & \colhead{I} & \colhead{J\tablenotemark{a}} & \colhead{H\tablenotemark{a}} & \colhead{K$_S$\tablenotemark{a}} & \colhead{Spec. Type} & \colhead{IR excess\tablenotemark{b}} &\colhead{RV} &\colhead{EW(Li)} & \colhead{Note} \\
& \colhead{(J2000)} & \colhead{(J2000)} & \colhead{(mag) }& \colhead{(mag)} & \colhead{(mag)} & \colhead{(mag)} & \colhead{(mag)} & \colhead{(mag)} & \colhead{(mag)} & & & \colhead{(\kms)} &  \colhead{(\AA)} &}

\startdata

  1&  83.24890&  -6.11734&     \nodata&     \nodata&     \nodata&     \nodata&      5.52&      4.30&      3.93&M8.0$\pm$2.0& N&\nodata&\nodata&g\\
  2&  83.26690&  -6.00261&     15.31&     13.80&     12.99&     12.22&     11.00&     10.27&     10.08&G6.0$\pm$2.0& N&\nodata&\nodata&g\\
  3&  83.31976&  -6.40979&     11.04&     10.45&     10.12&      9.81&      9.42&      9.12&      9.03&F9.0$\pm$2.0& N&\nodata&\nodata& f\\
  4&  83.33801&  -6.41991&     12.28&     10.72&      9.89&      9.13&      7.97&      7.19&      6.97&K3.0$\pm$3.0& N&\nodata&\nodata&g\\
  5&  83.36604&  -6.15616&     13.06&     11.90&     11.22&     10.56&      9.56&      8.97&      8.85&G2.0$\pm$2.0& N& 77.0& 0.0& def\\
  6&  83.40304&  -6.38011&     10.10&      9.46&     \nodata&      8.79&      8.28&      8.02&      7.96&G0.0$\pm$1.5& N& 23.5& 0.0& def\\
  7&  83.44242&  -6.46825&     13.27&     11.81&     10.99&     10.23&      9.03&      8.39&      8.18&K0.0$\pm$3.0& N&\nodata&\nodata&g\\
  8&  83.48413&  -6.31374&     14.46&     13.21&     12.51&     11.84&     10.81&     10.23&     10.06&G3.0$\pm$2.0& N&-18.5& 0.0& de\\
  9&  83.51110&  -6.09196&     14.77&     12.88&     11.75&     10.68&      9.07&      8.13&      7.91&F8.2$\pm$2.0& N&138.0& 0.0& de\\
 10&  83.54718&  -6.52211&     11.81&     10.87&     10.38&      9.93&      9.23&      8.80&      8.72&G5.0$\pm$2.0& N&\nodata&\nodata& f\\

\enddata
\label{tab:nonmember}

 \tablenotetext{a} {\scriptsize J,H,K$_S$ photometry is from 2MASS}
 \tablenotetext{b} {\scriptsize Criteria for IR excess are defined in Gutermuth et al. (2009) and Megeath et al. (2012).}
 \tablenotetext{c} {\scriptsize From Magellan/MIKE echelle data.}
 \tablenotetext{d} {\scriptsize Rejected based on radial velocity incompatible with the velocity of the cloud.}
 \tablenotetext{e} {\scriptsize Rejected based on lack of Lithium absorption.}
 \tablenotetext{f} {\scriptsize Rejected based on large proper motion.}
 \tablenotetext{g} {\scriptsize Rejected based on position on the color-magnitude diagram.}

\end{deluxetable*}

In this section, we discuss how we determine membership and rule out background/foreground contamination as well as the results of
optical spectroscopy and photometry of intermediate mass stars. Table~\ref{tab:highmass} lists the RA, Dec, photometric BVRI$_{\rm C}$ 
magnitudes and spectral types of targets that we classify as members of L1641 (with some uncertain members in the F \& G range, see 
discussion in~\ref{sec:member}). Table~\ref{tab:nonmember} lists objects that satisfy our selection criteria in \S~\ref{sec:selection} and 
observed with OSMOS targets, but ruled out as nonmembers.

\subsection{Spectroscopy}
We use SPTCLASS to spectral-type our targets \citep{Hernandez04}, a semi-automatic spectral-typing program. It uses empirical relations
of spectral type and equivalent widths to classify stars. It has three schemes optimized for different mass ranges (K5 or later, late F to early
K and F5 or earlier), which use different sets of lines. The user has to manually choose the best scheme for each star based on the
prominent features in the spectrum and the consistency of several indicators. While SPTCLASS is insensitive to reddening and S/N of the
spectra (as long as we have enough S/N to estimate the spectral indices), it does not take into account the effect of the hot continuum 
emission produced by the accretion shocks. This continuum emission makes the photospheric absorption lines appear weaker. 
SPTCLASS generally assigns
an earlier spectral type to veiled stars than their and therefore the SPTCLASS outputs should be considered as the earliest spectral type
limits.  While SPTCLASS utilizes spectral lines from 4000\AA~to 9000\AA, our OSMOS observations only cover a small spectral range from
3100\AA~to 5950\AA. As a result, SPTCLASS was unable to classify some of the K type stars, which we classified by eye and assigned a
larger uncertainty. Spectra of B \& A stars are also rectified and classified by eye as a double-check, and the results are always similar to
the SPTCLASS output within the errors. Figure~\ref{fig:spec} shows some sample spectra.

The earliest star found in the L1641 is a B0.5 star right at the very northern edge of the field. There are three other B1 to B3 stars 
within -6.1$^\circ<$Dec $< -6.0^\circ$. The earliest star south of this range is a B4 star at (RA,Dec) = (5:42:21.3, -08:08:00), first identified 
by \citet{Racine68}. Among the 57 stars identified as members, there are 4 B0-B3 stars, 14 B4 - B9 stars, 9 A stars, 15 F stars, 5 G stars 
and 10 K stars. (Since we do not have radial velocity for our entire sample, we expect some of the A, F \& G stars listed here are line-of-sight
contamination. See~\ref{sec:radial}.)

 \begin{figure*}
	\begin{center}
		\includegraphics[scale=0.6]{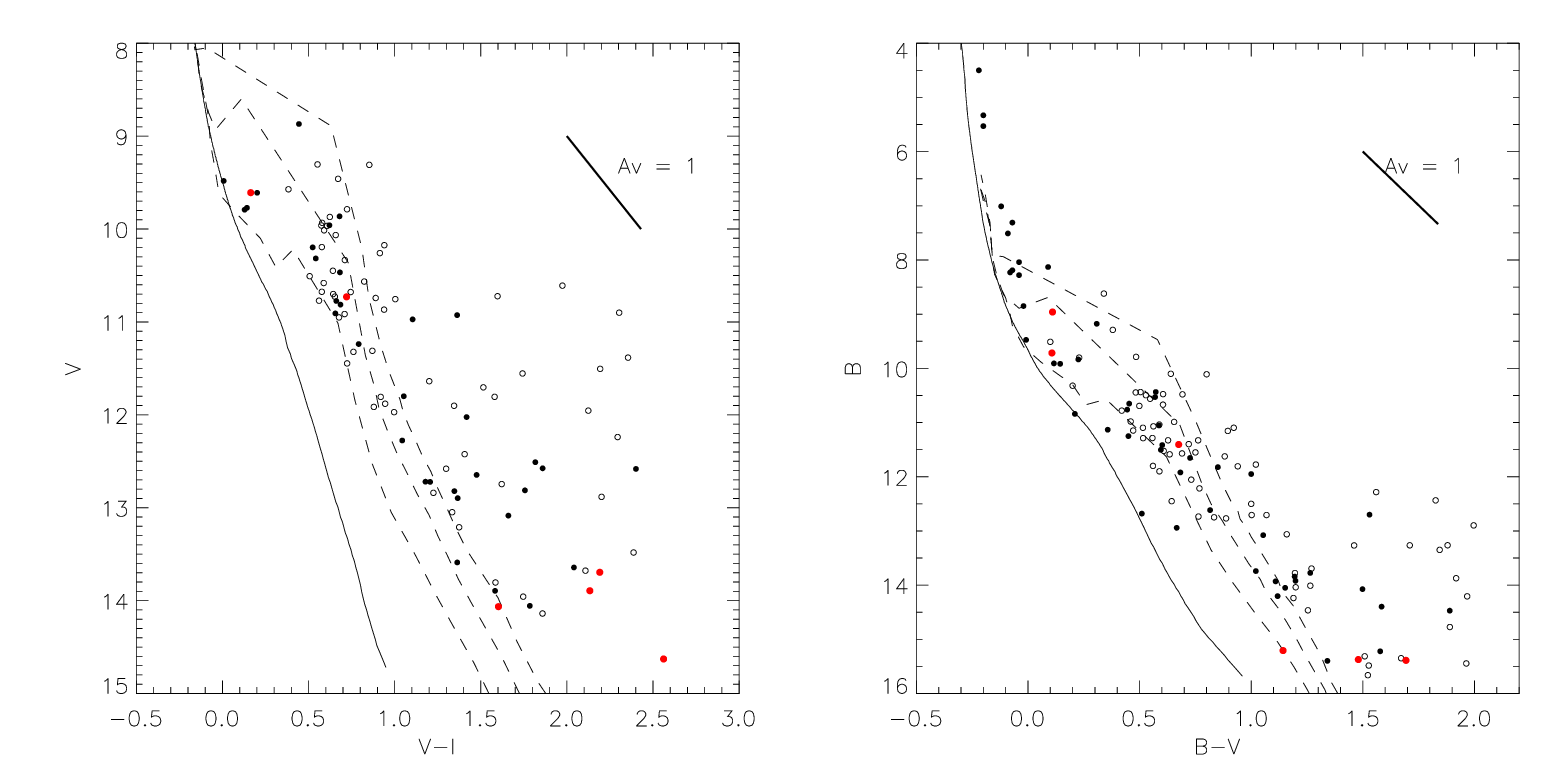}
		\includegraphics[scale=0.6]{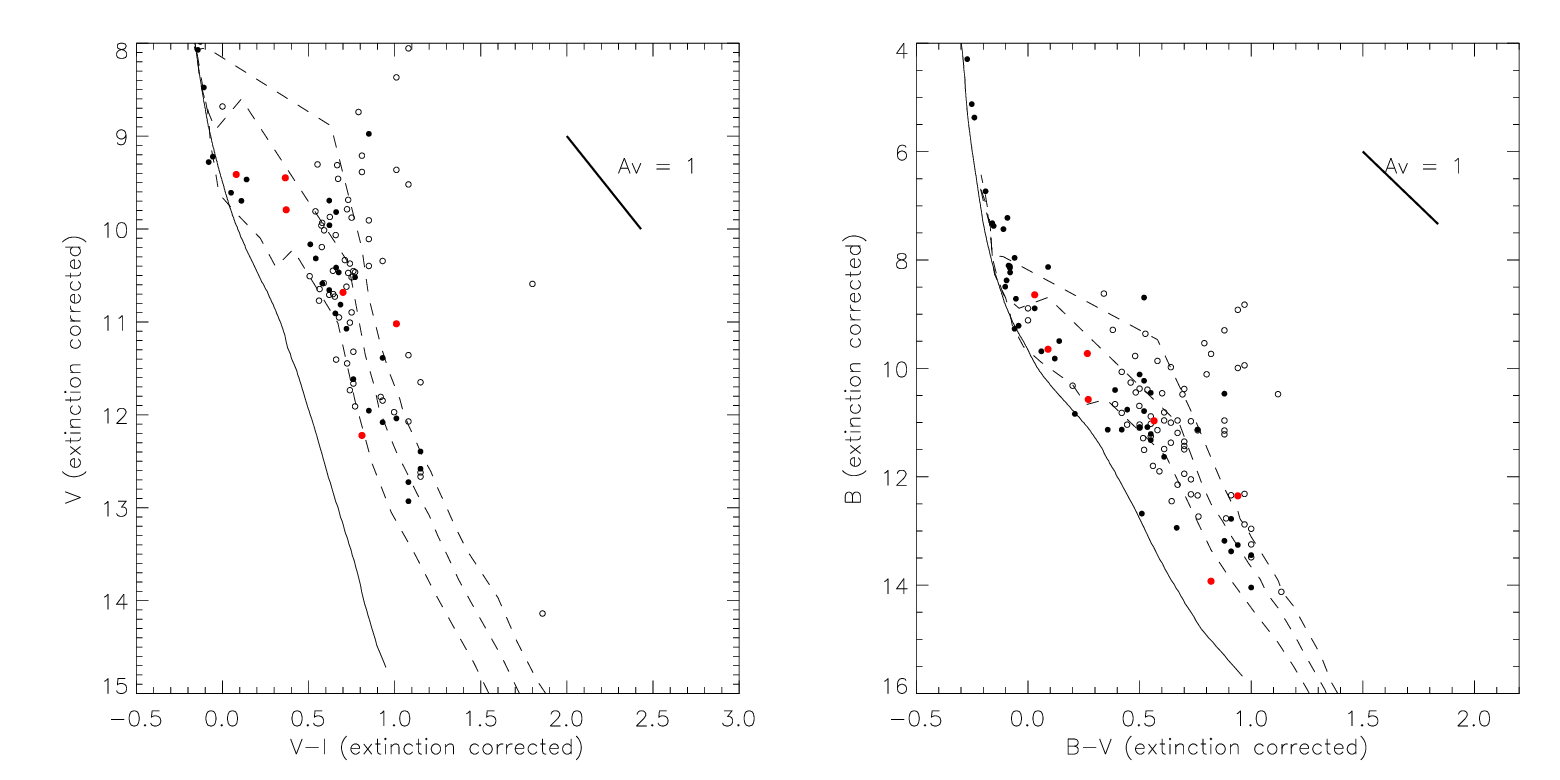}
	\end{center}
	\caption{V-I vs V and B-V vs B CMDs of high-to-intermediate mass stars in L1641 with photometry. The dashed lines are 2, 4, and 
	10 Myr \citet{Siess00} isochrones. The top panel shows CMDs without extinction correction; the bottom panel shows the same CMDs 
	after correcting for extinction, assuming intrinsic colors of Kenyon \& Hartmann (1995). The red symbols represent objects with 
	IR excess. The open circles represent non-members and solid circles represent members.}
	\label{fig:CMD}
\end{figure*}

\subsection{Photometry}
Figure~\ref{fig:CMD} shows the V-I vs. V and B-V vs. B color-magnitude diagram of all the members listed in Table~\ref{tab:highmass} that 
have magnitude information in the bands plotted.
For stars brighter than V$\sim$ 11 the OSMOS observations might be saturated and the Tycho-2 magnitudes are used. 
Note that the most massive members are only shown in the B vs B-V CMD because we do not have their I band magnitude. The top 
panels show the CMD without extinction correction. The red symbols are objects with IR excess and the black symbols are the non-IR
excess objects, as identified by \citet{Megeath12}. 
The bottom panels show the CMD with extinction correction, estimated by assuming intrinsic colors from \citet[][KH95]{KenyonHartmann95}
for stars earlier than M4 and \citet{Leggett92} for stars M4 and later and  a standard extinction law with R$_V$ of 3.1 
\citep{Cardelli89}. Overplotted are the ZAMS (solid) and 2, 4, and 10 Myr isochrones from \citet{Siess00}. The earliest type stars are
essentially on the main sequence whereas the later type stars are still above the main sequence. 

Stars that are identified as non-members are removed from this figure. The background giants generally have large extinctions and lie 
above the 2 Myr isochrone around V-I of 0.5 - 1 and B-V of 1 in the extinction corrected CMD. Foreground objects are generally later type
stars that are well-above the isochrones. 

\subsection{Membership}\label{sec:member}
We expect significant contamination by non-members in our sample given the distance of L1641 and its large area on the sky.
In paper I, we used indicators of youth such as Li absorption and H$\alpha$ emission in low-resolution optical spectra as well as the 
presence of IR-excess to determine membership for late type stars. In the intermediate mass range (A to G), confirming membership is 
more challenging. First of all, the disk fraction is lower for earlier type stars so we only find a small fraction of members with IR-excess. 
Secondly, intermediate mass stars have stronger H$\alpha$ absorption lines which can obscure H$\alpha$ emission from accretion. 
Finally, Li depletion timescales are longer for intermediate mass stars than for low mass stars. As demonstrated in \citet{Briceno97}, G 
type stars in Pleiades also show Li absorption, and the equivalent widths are hard to distinguish from that of pre-main sequence stars.  
As a result, we have to use additional kinematic information to help determine membership,
and combine this with statistical estimates of contamination from the Besan\c{c}on model of
stellar populations \citep{Robin03}as described in \S 3.4. 

 \begin{figure}
	\begin{center}
		\includegraphics[scale=0.7]{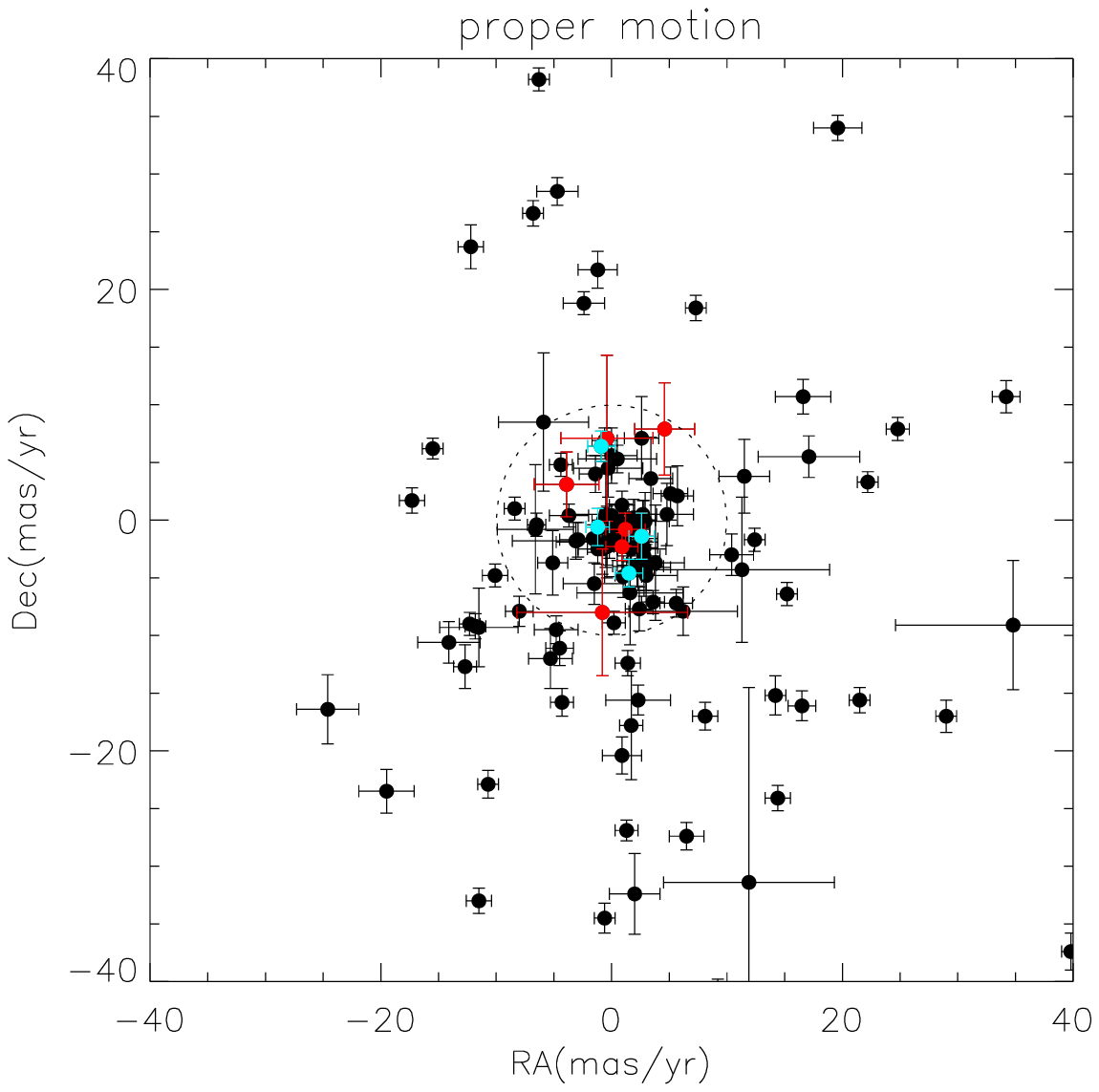}
	\end{center}
	\caption{Proper motion of our targets from the fourth U.S. Naval Observatory CCD Astrograph Catalog (UCAC4; \citealt{Zacharias12}).
	Stars with proper motions greater than 10 mas/year, or 20\kms, after the proper motion error is taken into account, are considered 
	foreground objects. The red and cyan symbols represent members confirmed through IR-excess and radial velocity with MIKE echelle 
	observations, respectively. The confirmed members all have small proper motions, well within the 10 mas/year cutoff applied.}
	\label{fig:proper}
\end{figure}

\subsubsection{Proper Motion}
One way to help eliminate foreground stars is through their proper motions. At 
the $\sim 414$~pc distance of Orion, we do not expect to detect large angular proper 
motions; stars with large proper motions are thus most likely foreground
objects. In Figure~\ref{fig:proper} we plot the proper motions of our 
OSMOS targets from the UCAC4 \citep{Zacharias12}. Most of our targets, as expected, show zero or very small proper motions. 
We fit Gaussians to the proper motion distributions in RA and Dec. The Gaussian fit to the proper motion distribution in RA peaks at 
0.2~mas/yr and has a width of  2.0~mas/yr; the distribution in Dec peaks at -2.0~mas/yr with a width of 2.7~mas/yr. 

We therefore apply a 3$\sigma$ cut of 10 mas/year, or 20 \kms~and stars outside the circle (after taking into account of the proper 
motion errors) are considered foreground objects. Note that while 90\% of the objects in this plot has errors of $<5$~mas/yr, a handful of 
objects have larger proper motion errors and are therefore compatible with zero proper motion. This will make our effort to eliminate 
foreground objects less effective.

 \begin{figure}
	\begin{center}
		\includegraphics[scale=0.7]{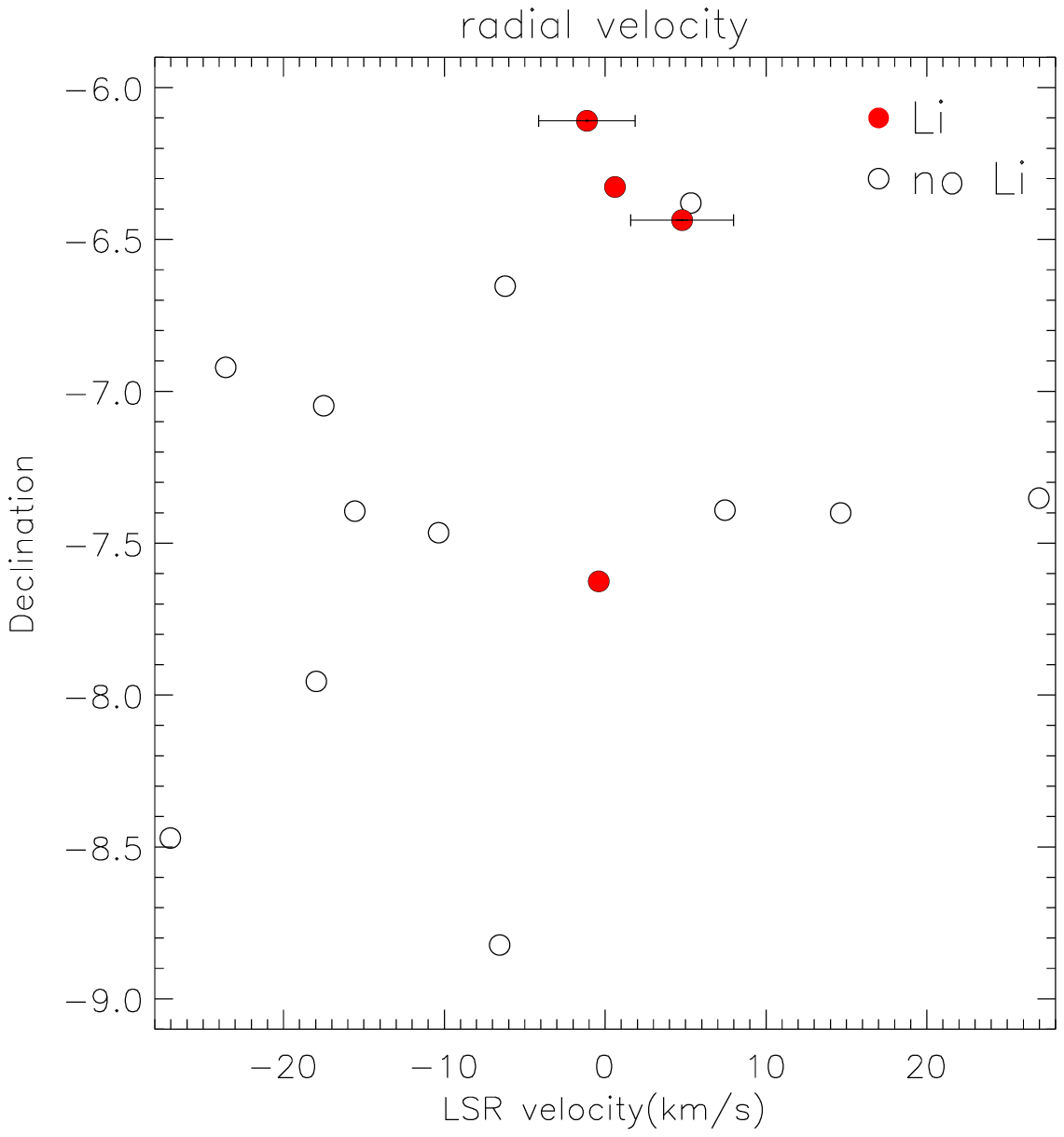}
	\end{center}
	\caption{LSR velocities and Li absorption data obtained from Echelle observations of 20 F \& G type stars in L1641. The red symbols 
	are stars with Li absorption above the level seen in Pleiades \citep{Jones96}. A member of L1641 should have velocities compatible 
	(within 5\kms) with the gas velocity ( 2 - 10 \kms, \citealt{Bally87}) and Li absorption. The typical errors in velocity for slow-rotating 
	stars are about 0.5\kms, comparable to the size of the points. The points with error bars overplotted are fast-rotators, where velocity 
	estimates are less certain.}
		\label{fig:rv_Li}
\end{figure}

\subsubsection{Radial Velocity}\label{sec:radial}
We observed 20 out of the 45 stars we classified as potential F \& G members of L1641 with the MIKE echelle spectrograph. The radial 
velocities of these 20 objects are listed in Table~\ref{tab:highmass} and~\ref{tab:nonmember} and shown in Figure~\ref{fig:rv_Li}. Members 
of L1641 should have LSR radial velocities compatible with the gas velocity (2 - 10 \kms, with a gradient along north-south direction, 
\citealt{Bally87}). Even though F \& G stars usually have very weak Li absorption, we are able to measure the equivalent widths at high 
spectral resolution. In order to be considered a member, a star should have velocities compatible with the gas (we chose a range of -3 - 15 
\kms, or within 5\kms of the gas velocity) and also Li absorption stronger than that observed in Pleiades \citep{Briceno97}. 
In Figure~\ref{fig:rv_Li}, we can see that most of the F\& G stars observed do not have the cloud velocities and can be ruled out easily. 
There are 4 members that are compatible with the cloud velocities and show Li absorption. There are also one or two stars that have the 
cloud velocity but do not show Li absorption. The typical error for radial velocity determination is about 0.5\kms, or the size of the dots. The 
larger error bars indicate fast rotating stars where it is more difficult to determine the radial velocity as accurately. However, since stellar 
rotation decreases with time \citep{Skumanich72}, fast rotation is also a sign that these stars are young and therefore they are considered 
members.

\begin{figure}
	\begin{center}
		\includegraphics[scale=0.6]{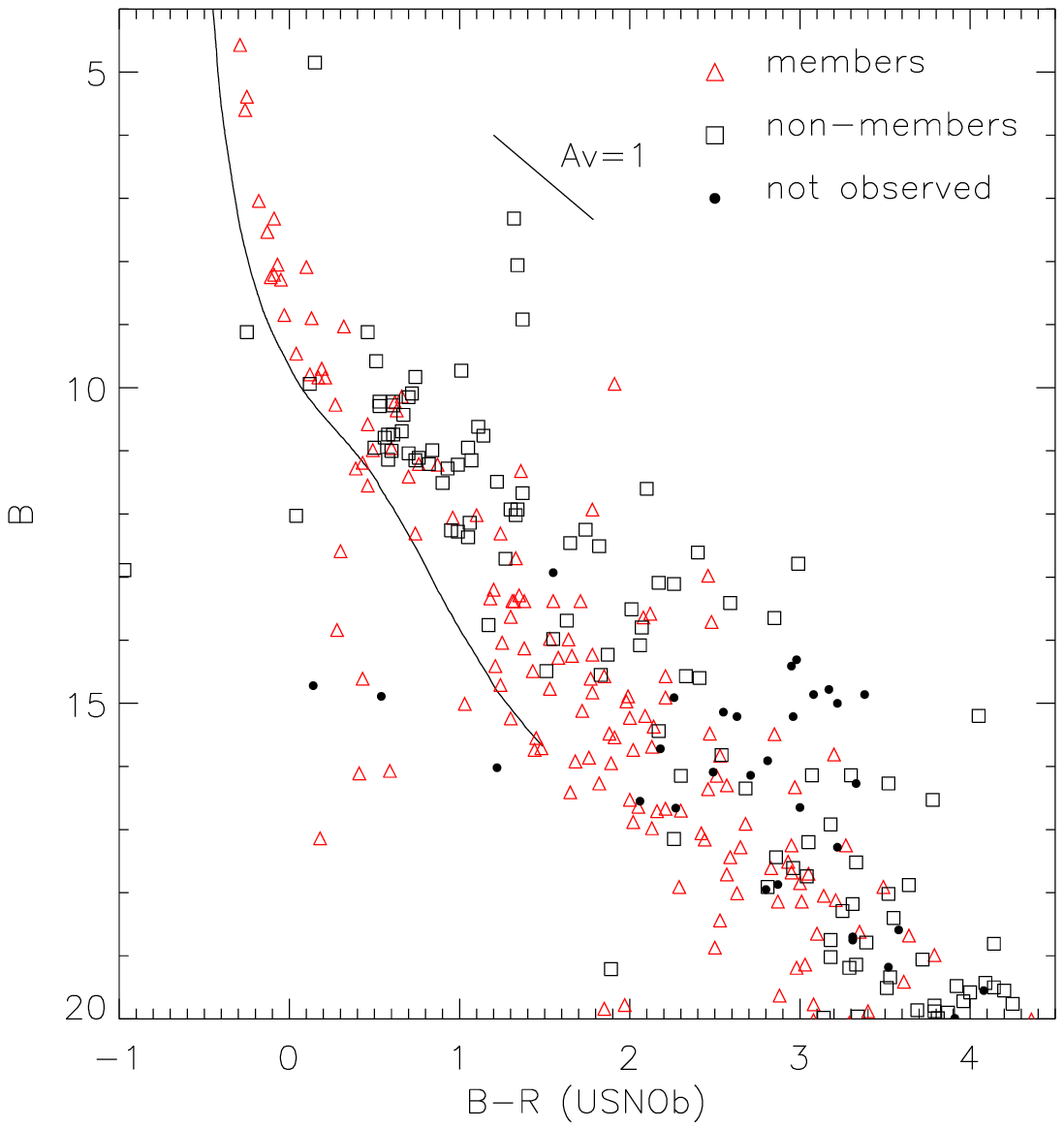}
	\end{center}
	\caption{USNOb B vs. B-R CMD of candidate intermediate mass stars selected through 2MASS. The red triangles are members 
	of L1641, the black boxes are non-members of L164 (see \S~\ref{sec:member} for how membership is determined). 
	 The black dots are objects that were not observed. Note that some of the spectral types and membership
	  information are obtained from Paper I. }
	  \label{fig:member}
\end{figure}

To summarize the result of the membership determination, 79 out the 136 stars in our intermediate mass sample are ruled out as 
non-members from kinematics and their position on the CMD. Only 4 out of 20, or 20\%, of the F \& G type stars observed with MIKE turn 
out to be members. In the following discussions, we will assume that only 20\% of the stars not observed with MIKE are members. 
However, as we do not know which stars in our sample are non-members,
we retain them in Table 1 and Figure~\ref{fig:member}. 

\subsection{Completeness, Contamination and Sample Definition}\label{sec:sample}

From here on, we combine the sample of high-to-intermediate mass stars characterized above and the confirmed members of L1641
cataloged in paper I for our analysis. The sample from paper I has a total of 864 members, including our Hectospec and IMACS targets
as well as the sample from \citet{Fang09} and reference therein. The spectral types range from F to M4, where the F \& G stars are mostly 
from the literature and not from our own observations. We assessed the completeness of the low-mass sample by comparing the number of 
IR-excess stars in the \citet{Megeath12} sample and our sample. If we limit ourselves to objects with \Av $<$ 2 for 1 to 0.1\Msun, then
our spectroscopic sample is about 90\% complete. Since we required Li absorption and/or IR-excess for membership, we expect very few
contaminants in the low-mass sample. We do note that the completeness is worse for the non-excess objects because we are not able to 
target them as efficiently. In particular, we are missing non-IR excess objects in the G \& K range that are too bright for 6.5m telescopes and
as well as the lowest-mass objects because it is more difficult to observe Li absorption in lower S/N spectra. To summarize, we are missing
members in the low-mass sample and the incompleteness depends somewhat on the spectral type. However, in \S~\ref{sec:comparison}, 
we will explain why this incompleteness does not affect our result as long as we use the lower limit of the number of low-mass stars.

Contamination rather than completeness is the problem for the higher mass objects, and identification of members is more difficult.
While we can rule out foreground dwarfs and background giants to some extent based on their position in the color-magnitude diagram, 
but we will still have non-members in the sample that cannot be ruled out simply by proper motions and/or for which we have no radial 
velocity measurements. To estimate the possible contamination, we use the Besan\c{c}on model \citep{Robin03} to help us determine 
the number and type of contaminants we would find along the line of sight in a 3$^\circ$ field toward the direction of L1641. 

We first consider the range of O and B stars. Since we are trying to test the IMF variation in 
this mass range, it is essential that we have a complete sample. This is achieved by our survey of the brightest 2MASS stars, regardless of
whether they have IR-excess. We note that because of step (3) in our target
selection (see \S~\ref{sec:selection}), highly embedded O or early B stars would not have been found. However, evolved O or early-B stars
would have blown away materials nearby in a manner similar to the B4 star near (RA, Dec) = (5:42:21.3, -08:08:00) and therefore would 
not have high extinction. (Note that there is another reflection nebula, powered by a potentially massive star further south, outside of the 
fields of our optical studies.) On the other hand, if there were unevolved, highly-extincted early-B stars, they would heat up the dust nearby
and be identified as bright IR-excess sources by the {\em Spitzer} survey. We examined the brightest IR-excess sources in the field and 
confirmed that they are mostly A stars, with one exception being an FU Ori star (V883 Ori), where the IR luminosity is dominated by
accretion. We also examined the protostars that have large bolometric luminosities. The most luminous protostar in L1641 is located at 
(RA, Dec) = (5:40:27.4,-7:27:30) in one of the high-extinction clumps. It has L $\sim$ 490 \Lsun \citep{Kryukova12}, and according to the 
Siess (2000) isochrones, has a mass less than 7 \Msun. Because of the short lifetime and relative small number of OB stars, we expect no
contamination in this mass range that is at the right distance to appear along the ZAMS at 414 pc. The Besan\c{c}on model predicts no OB 
stars in the field within 1 kpc. 

For A stars, we expect our survey to be complete to at least \Av = 2 since we targeted stars that are brighter than B = 15 and B-R $<$ 2.5 
in the USNOb catalog. Unfortunately, we do not have echelle observations of the A stars, and are therefore unable to constrain membership 
rate from 
radial velocities.  We are left with estimating the level 
of contamination is the Besan\c{c}on model, which predicts 3 foreground A 
stars above the region's isochrone and up to 5 background giants. The
actual number of background giants that make it into our survey depends on the extinction 
toward them and is therefore more difficult to constrain. 
In Table~\ref{tab:highmass} and~\ref{tab:nonmember}, we identified 3 A stars as non-members and 9 A stars as members. From the 
discussion above, we therefore conclude that out of the 9 stars considered members, up to 5 of them can actually be non-members. 
If we adopt a moderate extinction of \Av = 2, we expect 3 non-members.

We expect a large number of line of sight contamination for F \& G stars. In \S~\ref{sec:member}, we determined that only 20\% of the observed 
F \& G stars are actual members based on radial velocities. 
Our sample also suffers from incompleteness in the G star range 
because of the target selection criteria
in our intermediate-mass and low-mass samples tend to miss G stars that are below the 1\Msun~line in the J vs. J-H CMD.

As discussed in Paper I, our optical photometric and spectroscopic sample include stars in the Orion A cloud south 
of -6$^\circ$, which we now define as the L1641-all sample. However, within this sample, the northern end is a 
relatively denser region and has four early-B stars that likely denote the outer regions of the ONC. We therefore 
further divide the L1641-all sample into subsamples L1641-n (Dec=$-6.5^\circ$ - $-6^\circ$ and L1641-s 
(Dec$<-6.5^\circ$). Note that this definition is only for the purpose of analysis in this paper and does not coincide 
with the small clusters L1641-North (RA$\sim 84.1^\circ$, Dec$\sim-6.3^\circ$) and L1641-South 
(RA$\sim85.7^\circ$, Dec$\sim -8.2^\circ$).

\subsection{HR diagram}

We use an HR diagram to determine the ages and masses of L1641 members.  We use the effective temperature scale from 
KH95 for stars earlier than M4 and effective temperatures from \citet{Luhman03} for stars M4 and later. If the I$_{\rm C}$ band 
magnitude is available for a star, we use the following equation to determine the bolometric luminosity:
\begin{equation}
\log (L/L\odot) = 0.4~[M_{\rm bol,\odot} - (I_{\rm C} - A_{I_{\rm C}}) - DM + BC_{I_{\rm C}}]
\end{equation}
where the bolometric magnitude of the sun $M_{\rm bol,\odot}$ is set to 4.75 and $A_{I_{\rm C}}$ is the extinction at $I_{\rm C}$ 
determined by dereddening the the observed $V-I_{\rm C}$ color. 

If the I$_{\rm C}$ band magnitude is not available (only the brightest stars), we use a similar equation using the V band magnitude
to determine the bolometric luminosity:
\begin{equation}
\log (L/L\odot) = 0.4~[M_{\rm bol,\odot} - (V - A_V) - DM + BC_V]. 
\end{equation}
where \Av~is the extinction at V determined by dereddening the B-V color.
We use bolometric corrections for the corresponding magnitude band from KH95 for stars earlier than M4 and 
\citet{Bessell91} for stars M4 and later. We assume that L1641 is at the same distance as the ONC (414 pc; \citealt{Menten07}), 
corresponding to distance modulus DM = m-M = 5$\log_{10}({\rm d}/10{\rm pc})$= 8.09. To determine the extinction, we use the V-I$_C$ 
colors when both bands are available. 
For the brightest stars where the data is taken from the Tycho2 catalog, the B-V color is used to estimate the extinction. The intrinsic 
colors are from KH95 for stars earlier than M4 and \citet{Leggett92} for stars M4 and later. We then calculate the extinction assuming 
a standard extinction law with R$_V$ of 3.1 \citep{Cardelli89}. 

 \begin{figure*}
	\begin{center}
		\includegraphics[scale=1]{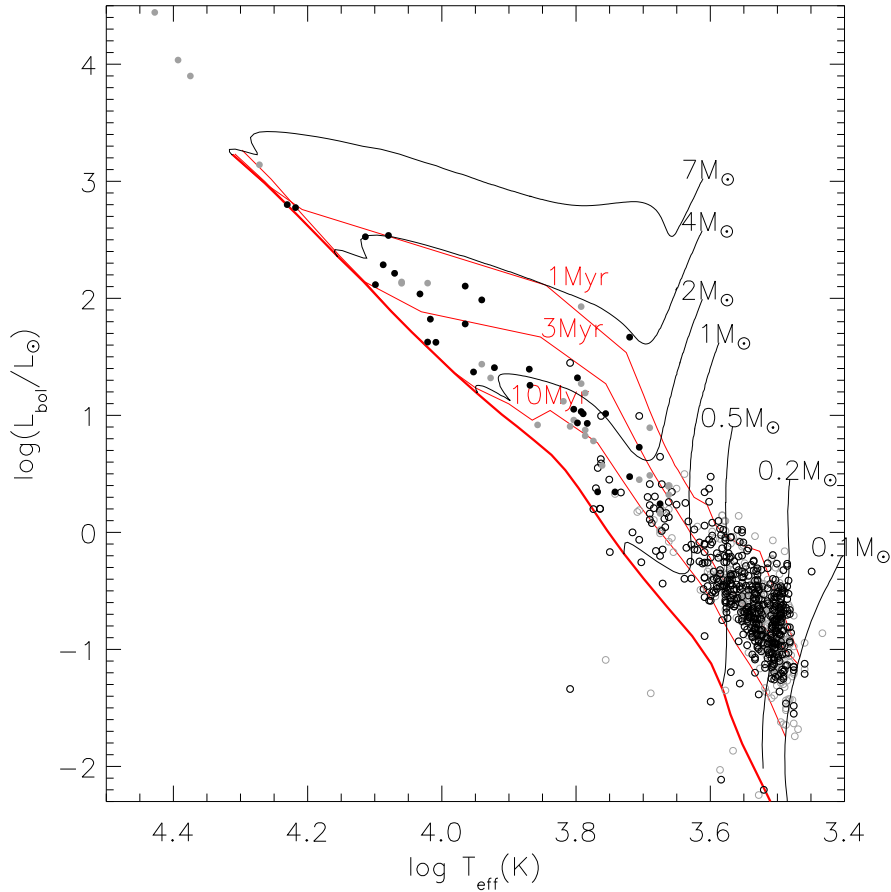}
	\end{center}
	\caption{HR diagram of members of L1641. The black symbols represents stars in L1641-s and the lighter 
	grey symbols are stars in L1641-n (see \S~\ref{sec:sample} for the definition of the populations). The 1, 3 and 
	10 Myr \citet{Siess00} isochrones (Z=0.02, no overshoot) 
	overplotted in red. The thick red line corresponds to the ``early main-sequence", defined as the time when the star settles on 
	the main sequence after the CN cycle has reached its equilibrium. In black are the evolutionary tracks for 
	masses from 0.1 to 7\Msun.}
	\label{fig:HR}
\end{figure*}

Figure~\ref{fig:HR} shows the HR diagram. The black symbols represents stars in L1641-s and the lighter grey 
symbols are stars in L1641-n  (see \S~\ref{sec:sample} for the definition of the populations). The 1, 3 and 10Myr 
\citet{Siess00} isochrones (Z=0.02, no overshoot) overplotted in red. 
The thick red line corresponds to the ``early main-sequence", defined as the time when the star settles on the main sequence after the CN 
cycle has reached its equilibrium. In black are the evolutionary tracks for masses from 0.1 to 7\Msun. We can estimate the ages and 
masses from the isochrones. We first select data points on the Siess tracks with ages less than 100 Myrs and pre-main sequence or main 
sequence evolutionary phase. We then use the IDL programs {\em TRIANGULATE} and {\em TRIGRID} to interpolate between the 
evolutionary tracks to obtain the ages and masses. Objects that fall outside of the track area are not assigned masses or ages. 

\subsection{Age and Mass distribution}\label{sec:age_mass}
\begin{figure}
	\begin{center}
		\includegraphics[scale=0.6]{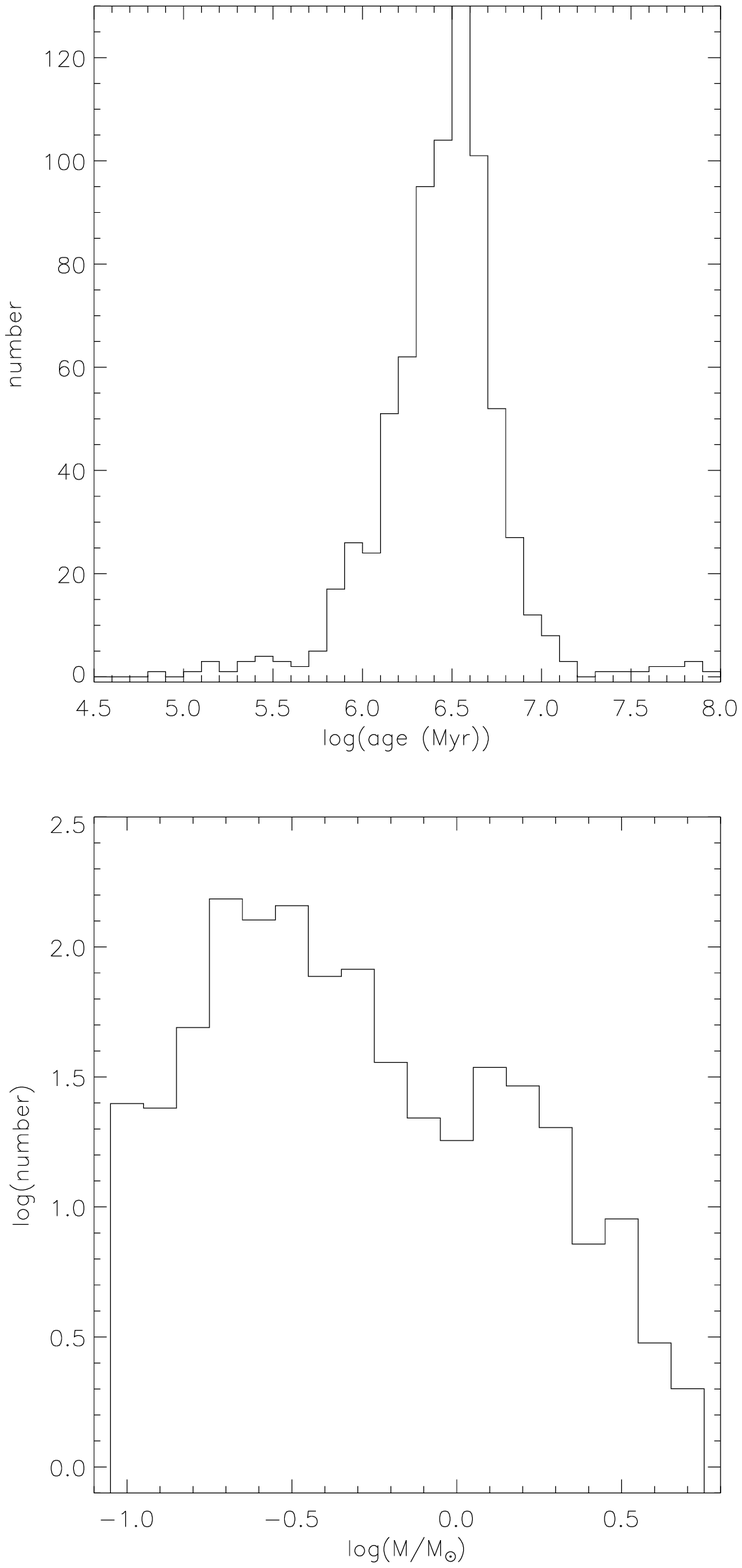}
	\end{center}
	\caption{Top panel: Age distribution of objects with masses below 1\Msun as inferred from the \citet{Siess00} 
	 isochrones. We only consider objects with masses below 1\Msun in the age estimates because the birthline 
	 effect tend to make the highmass objects appear older and the low-mass objects give a 
	 more reliable age estimate of the population. Bottom panel: Mass distribution of all the members.}
	\label{fig:age_mass}
\end{figure}

The top panel of Figure~\ref{fig:age_mass} shows the age distribution of the members in L1641 estimated from the HR diagram. Only 
objects with masses below 1\Msun~are considered in the age estimates because the birthline effect tend to make the highmass objects 
appear older (see \eg \citealt{Hartmann03}). Both the median and the mean of the age distribution is around 6.5, with a standard deviation 
of 0.3 dex. Therefore, the majority of the low-mass members in L1641 are around 10$^{6.5}$ years, consistent with the ages of the ONC. 
This is the same as the result found in Paper I where we used a more direct age estimate from the V vs. V-I CMD. This ensures that 
L1641 is still young enough that any early B star, if ever present, would still be on the main sequence and seen in our survey. There is also
no evidence of SN bubbles from O stars. 

The bottom panel of Figure~\ref{fig:age_mass} shows the mass distribution of L1641 members in log-log space. Each of the F \& G 
unconfirmed members is considered 0.2 star in the histogram. Note that the mass histogram presented here is not extinction limited
and we did not correct for completeness. The dip around 1 \Msun~corresponds to the gap between the intermediate-mass sample and the
low-mass sample. This figure demonstrates that because of incompleteness and extinction issues, our sample does not allow us to obtain
the upper-mass IMF by simply fitting a power-low. 

\section{Comparing L1641 to the ONC}\label{sec:comparison}

We note that the selection criteria for our spectroscopic observations and when a star in our spectroscopic survey is considered a member  
is between Paper I and this work and the criteria are spectral type dependent. This is mainly because the youth indicators themselves are 
spectral-type dependent. As a result, the completeness and contamination rate is also spectral-type dependent. As a result, our sample is 
not a clean extinction-limited complete sample and we cannot test the form of the IMF via the distribution of spectral types and K-band 
luminosity. We can only compare the ratio of high-to-low mass objects.

 \begin{figure}
	\begin{center}
		\includegraphics[scale=0.5]{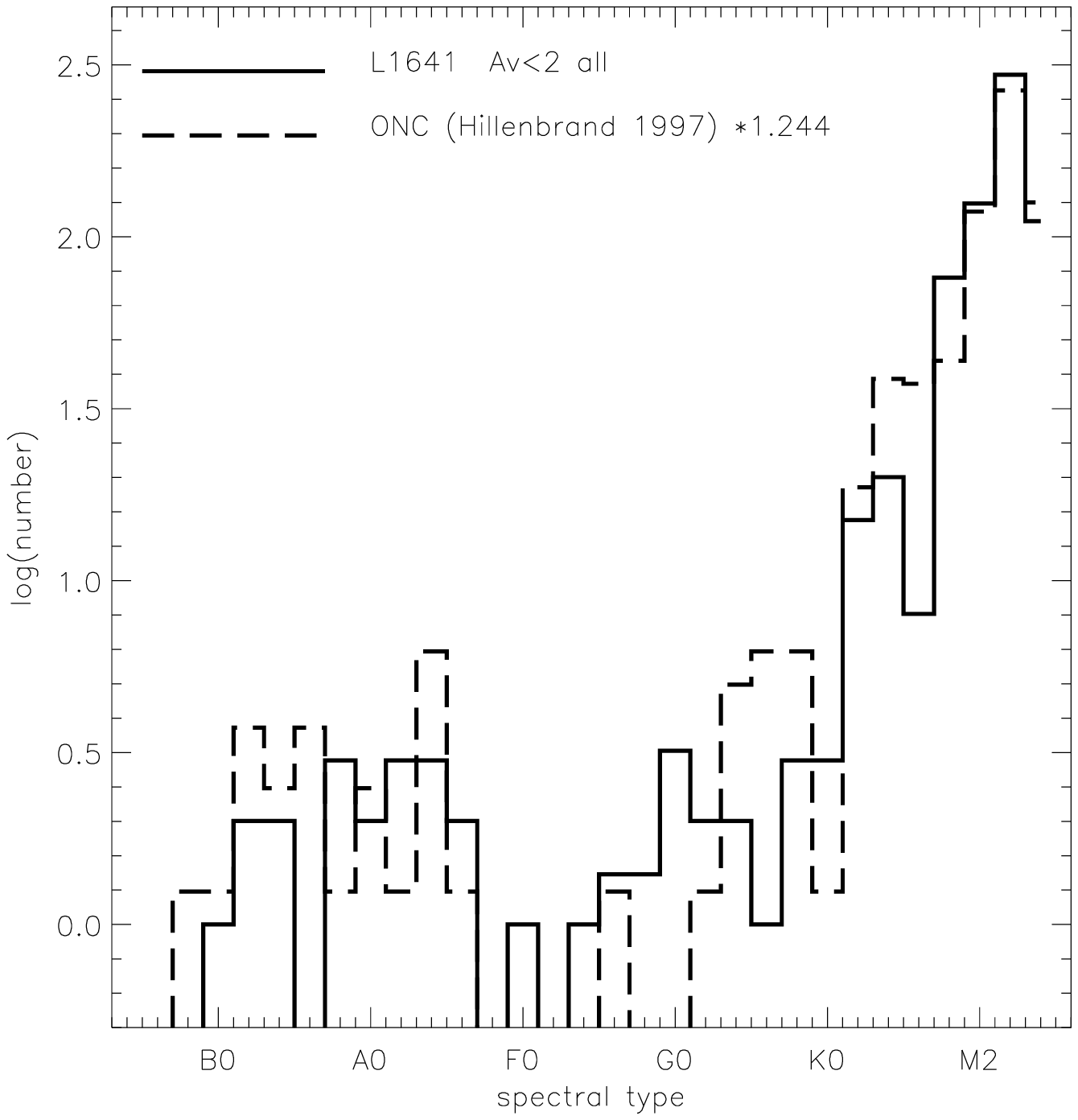}
		\includegraphics[scale=0.5]{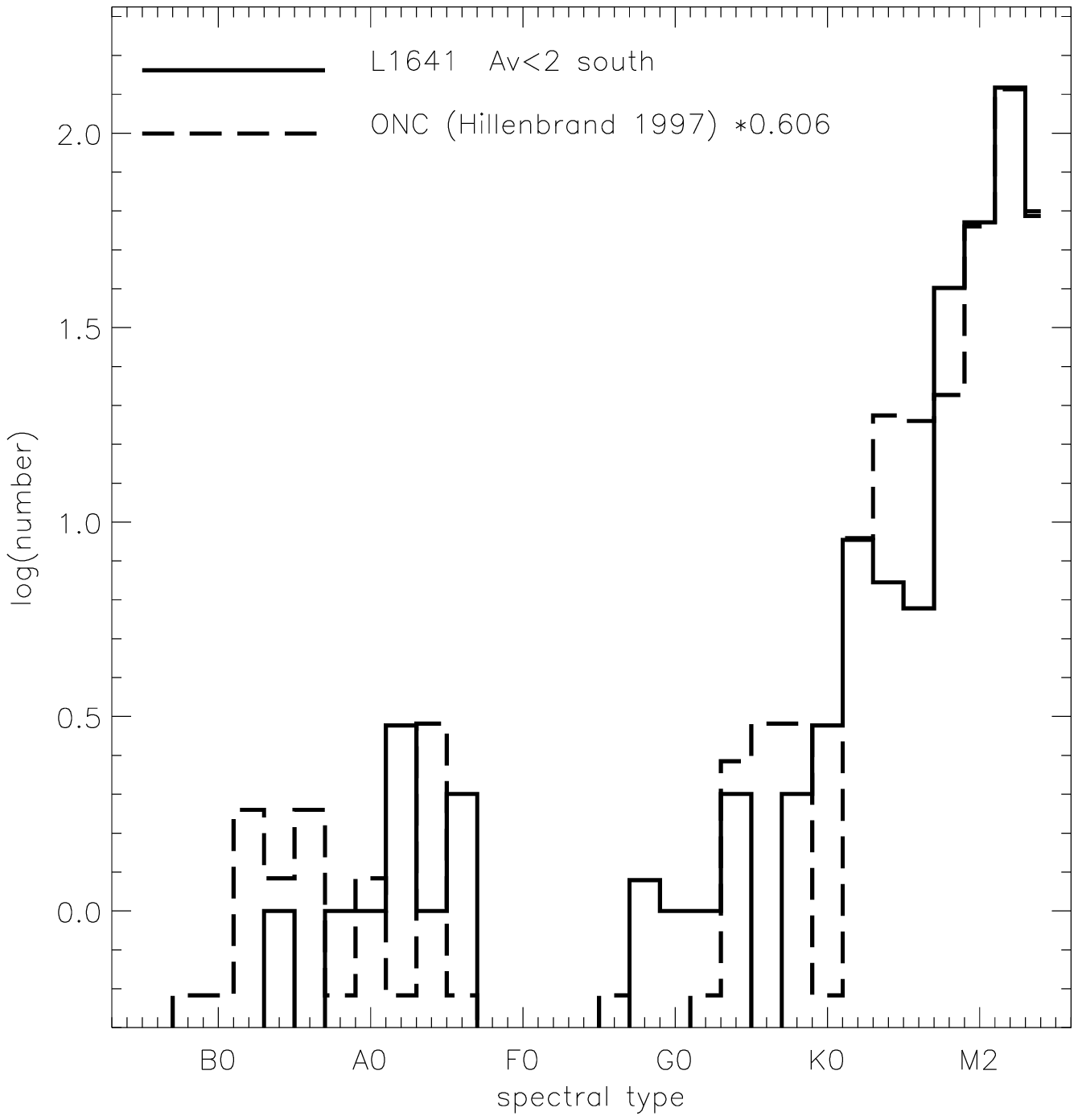}
	\end{center}
	\caption{Top: Histogram of spectral type distribution of the extinction limited sample (\Av $\le 2$) in L1641-all (solid line) and 
	in the ONC
	 (dashed line). Bottom: Same as top figure, but only considering stars in L1641-s. The ONC population is scaled to match the
	 size of the L1641 population, and the bins with only one star gives a negative value after scaling by a factor less than 1. }
	\label{fig:spt_dist}
\end{figure}

\subsection{Comparing L1641 and ONC spectral type distributions}
The top panel of Figure~\ref{fig:spt_dist} shows the spectral type distribution of the extinction limited sample (\Av $\le 2$) in L1641 and in
 the ONC. The solid line shows the spectral type distribution of L1641 members with \Av $\le 2$.  We first remove the objects that are 
 identified as non-members, and then we scale the F \& G stars whose membership status is uncertain by 20\%, based on the MIKE 
 observations indicating that only 20\% of them are actual members. 
 The dashed line shows the spectral type distribution of ONC members from H97 with \Av 
$\le 2$ and membership probability $>$ 70\%, scaled to the same number of stars as L1641. 

The bottom panel is the same, but shows only stars in L1641-s. We can see that L1641, especially the southern part is
 deficient in stars earlier than B4 compared to the ONC. We also note that L1641 has a smaller number of late G and early K stars 
 compared to the ONC. This is because we are missing some late G and early K stars from our target selection 
 (see \S~\ref{sec:selection}). Other than the missing members in this range, the low-mass end of the distribution look fairly similar. 

In Paper I, we found that the IMF of L1641-s is inconsistent with the standard IMF models. Compared to both 
\citet{Chabrier05} and \citet{Kroupa01} IMFs, the L1641 is deficient in O and early B stars to a 3-4$\sigma$ significance level. Here we 
discuss whether the ratio of high mass to low mass stars is different in L1641 and the ONC. However, instead of comparing the derived 
masses, a model dependent quantity, we choose to compare the directly-observable spectral types.

We then use the Fisher's exact test \citep{Fisher25} to find the significance level of the ratios between high-mass and low-mass stars in 
different regions of 
L1641 and the ONC. Since we are interested in whether the ONC has a higher frequency of high-mass stars, we use the one-sided test 
and do not compute the P value when the ONC has a lower frequency in the high-mass bin. The results are summarized in Table~\ref
{tab:spt_Fisher}. In general, when the entire L1641 population or the northern region is considered, the frequency of high-mass stars is 
consistent to that in the ONC. In the L1641-s region, there is some evidence that the high-mass IMF could be different, especially
when we consider the ratio of (O to B3 stars) to (B4 to M4 stars), where the significance level is 0.092. The difference in the high-mass IMF
is suggestive but not conclusive, mainly due to the small number of early-type stars in both samples. 

\begin{deluxetable*}{c|c|c|c|c|c}
\small
\tablecaption{Number of Stars in Spectral Type Bins}

\tablehead{\colhead{Region}  & \colhead{O stars} & \colhead{B0-B3} & \colhead{B4-B9} & \colhead{A stars} & \colhead{F0-M4}}

\startdata

ONC (H97)\tablenotemark{a}					&	2	&	4		&	7	&	7		&	437		\\	
\hline
L164-all1\tablenotemark{b}						&	0	&	4		&	6	&	8		& 	451		\\
\hline						
L1641-n\tablenotemark{b}	&	0	&	4		&	3	&	2		&	238		\\
\hline
L1641-s\tablenotemark{b}			&	0	&	0		&	3	&	6		&	213		\\

\enddata	
\tablenotetext{a} {Only stars with \Av$\le2$ and membership probability $>70$\% are considered.}
\tablenotetext{b} {Only stars with \Av$\le2$ are considered.}
\label{tab:spt_list}
\end{deluxetable*}
\begin{deluxetable*}{c|c|c|c}
\small
\tablecaption{Significance Level of Fisher's Exact Test on the Ratio of High-mass to Low-mass Stars in the ONC and L1641}

\tablehead{ \colhead{Compared Regions}  & \colhead{O stars/B0-M4} & \colhead{O-B3/B4-M4} & \colhead{O-B9/A0-M4} }

\startdata
ONC\tablenotemark{a} \& L1641-all\tablenotemark{b}	&	0.243	&	0.360	&	0.314	\\

\hline						
ONC\tablenotemark{a} \& L1641n\tablenotemark{b}	&	0.421	&	\nodata	&	0.599	\\
\hline
ONC\tablenotemark{a} \& L1641s\tablenotemark{b} 		&	0.452	&	0.092	&	0.176	\\

\enddata	
\tablenotetext{a} {Only stars with \Av$\le2$ and membership probability $>70$\% are considered. Data taken from H97.}
\tablenotetext{b} {Only stars with \Av$\le2$ are considered.}

\label{tab:spt_Fisher}
\end{deluxetable*}
\begin{deluxetable*}{c|c|c|c|c}
\small
\tablecaption{Number of Stars in K$_s$-magnitude Bins}

\tablehead{\colhead{Region}  & \colhead{$K_s < 7$} & \colhead{$7 < K_s < 8$} & \colhead{$8<K_s<9$} & \colhead{$9<K_s<12$}}
\startdata

ONC (\citet{Muench02})			&	8	&	10	&	29		&	397		\\	
\hline
L1641-all						&	4	&	10	&	23		&	518		\\
\hline						
L1641-n						&	3	&	2	&	6		&	170		\\
\hline
L1641-s						&	1	&	8	&	17		&	348		\\

\enddata	
\label{tab:KLF_list}
\end{deluxetable*}
\begin{deluxetable*}{c|c|c|c}
\small
\tablecaption{Significance Level of Fisher's Exact Test on the Ratio of High-mass to Low-mass Stars in the ONC and L1641}

\tablehead{ \colhead{Compared Regions} & \colhead{$K_s<7~/~7<K_s<12$} & \colhead{$K_s<8~/~8<K_s<12$} & \colhead{$K_s<9~/~9<K_s<12$} }

\startdata
ONC\tablenotemark{a} \& L1641-all	 & 0.103	&	0.118	&	0.018		\\

\hline						
ONC\tablenotemark{a} \& L1641-n	& 0.600	&	0.302	&	0.050		\\
\hline
ONC\tablenotemark{a} \& L1641-s	& 0.034	&	0.131	&	0.044		\\

\enddata	
\tablenotetext{a} {Data taken from \citet{Muench02}.}

\label{tab:KLF_Fisher}
\end{deluxetable*}

 \begin{figure}
	\begin{center}
		\includegraphics[scale=0.6]{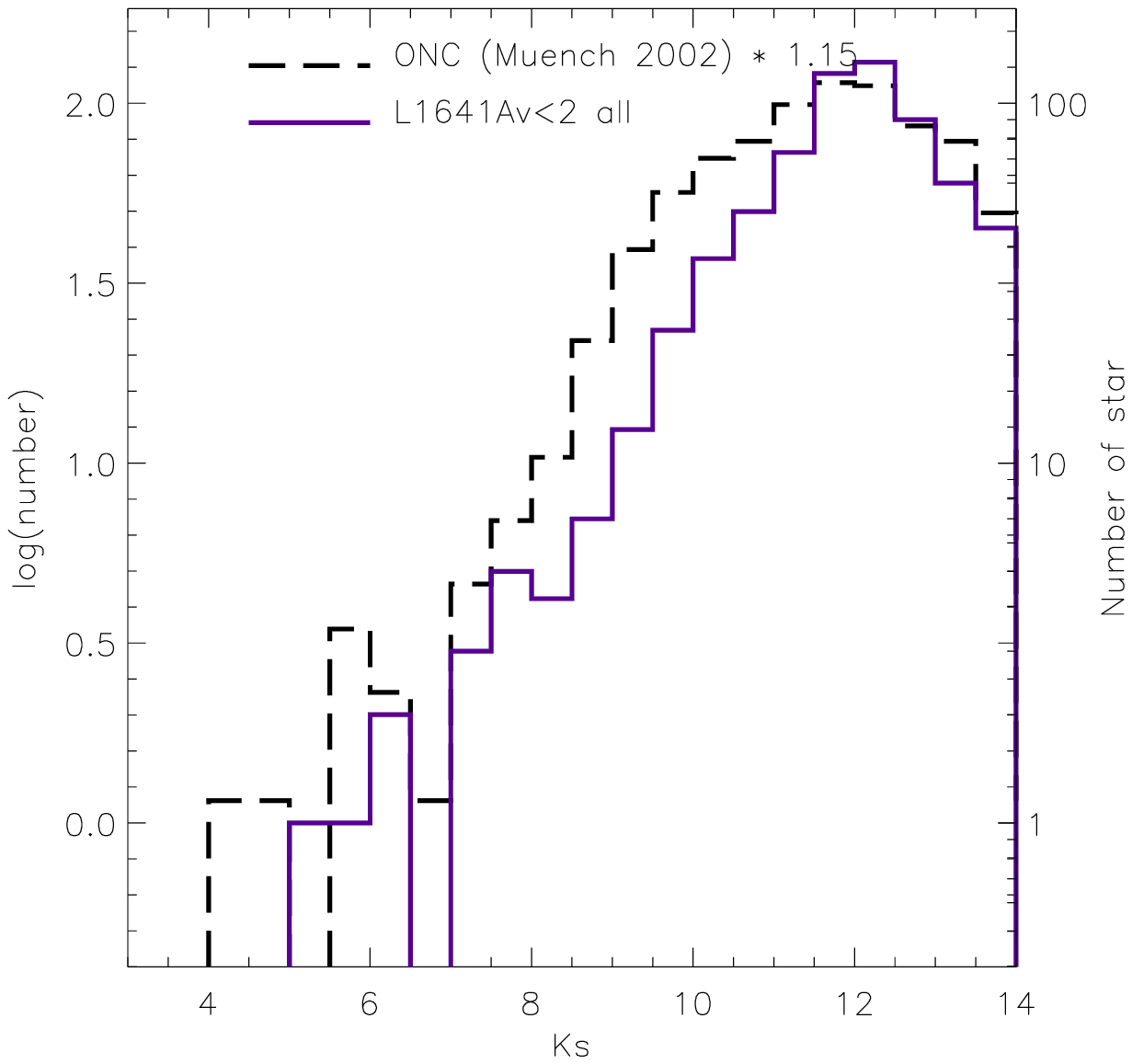}
		\includegraphics[scale=0.6]{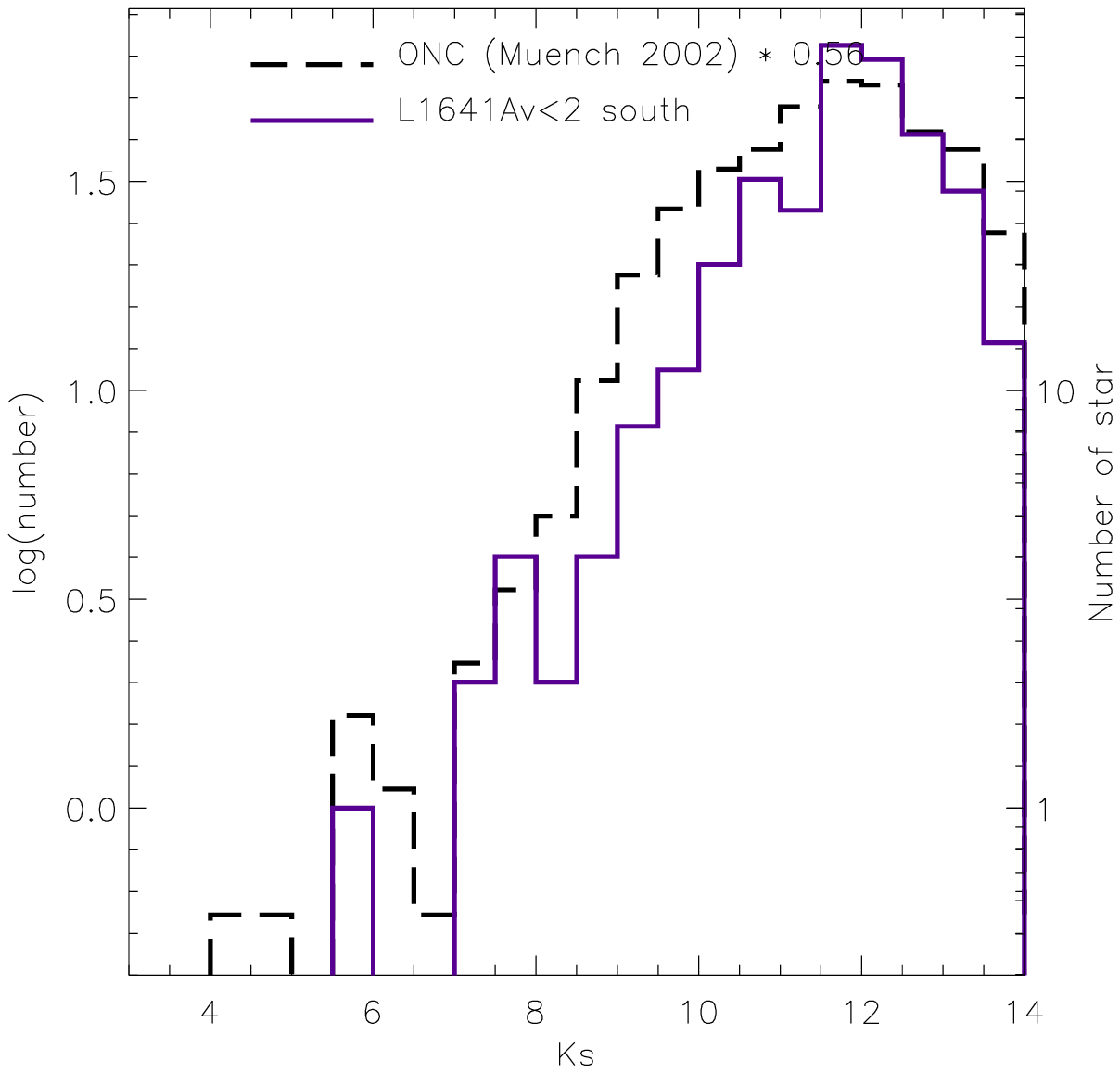}
	\end{center}
	\caption{K-band luminosity function (KLF) of extinction limited L1641 sample. Top: objects in L1641-all. Bottom:
	only objects in L1641-s. In both figures, the black long dashed lines represent the KLF of the Trapezium from
	\citep{Muench02}. The solid purple lines represent the KLF of objects in L1641.}
	\label{fig:KLF}
\end{figure}

\subsection{Comparing the L1641 and ONC K-band luminosity functions }


Next we compare the K-band luminosity function (KLF) of L1641 to that of the central regions of the ONC (the Trapezium cluster) 
\citep{Muench02}.

Table~\ref{tab:KLF_list} lists the number of observed stars in the Trapezium and confirmed members of L1641 in each magnitude bin, 
including objects that have extinction higher than \Av~=~2 from our optical survey as well as objects that are too extincted in the optical 
and are identified by IR-excess only. The study of \citet{Muench02} covered the central 5\arcmin $\times$ 5\arcmin of the Trapezium and it
is safe to assume that in the magnitude range of our interest (K$_{\rm s}$~brighter than12), the line-of-sight contamination is very small 
and can be ignored. In L1641, since the field is much larger and the density of stars much lower, the contaminants outnumbers the 
members and making an off-field correction can lead to large errors. We therefore obtain the number of all known members by adding the 
spectrally-confirmed members to the highly-extincted IR-excess members. The number of stars in each magnitude bin has varying 
completeness and is biased against highly-extincted low-mass objects that do not show IR-excess.

Figure~\ref{fig:KLF} shows the K-band luminosity function of extinction limited (\Av$\le 2$) L1641 sample. The top panel shows objects in 
L1641-all and the bottom panel shows only objects in L1641-s. 

We use the one-sided Fisher's exact test to compare the ratios of stars above and below the threshold magnitude. The hypothesis we want 
to test is that the L1641 has a lower proportion of stars in the high-mass bin compared to the Trapezium. We use cutoff magnitudes of \
K$_s$=7 and 8 to ensure that we are not missing members in the high-mass (bright) bin. Table~\ref{tab:KLF_Fisher} lists the significance 
levels of the one-sided Fisher's test. The Trapezium population has a higher fraction of high-mass stars compared to the L1641 
population as a whole even though the difference is not very significant. The L1641-s population has a significantly lower fraction 
of high-mass stars compared to the Trapezium, with a P value of 3\%. 

The KLF has the advantage of being minimally affected by extinction (the extinction in K band is about one tenth of the extinction in the V 
band). The KLF can also be affected by disk-excess. \citep{Muench02} constructed the K-band IR excess distribution function for the 
Trapezium, where the K band excess peaks near 0.2 with a mean of 0.4 mag. Even though disk-excess can change the shape of the KLF, 
this excess does not significantly affect our number counts in Table~\ref{tab:KLF_list} because of our coarse binning as well as the fact that 
most of the brightest objects are diskless. In addition, since L1641 and the ONC has similar ages, it is reasonable to assume the disk excess 
would have similar effects on the L1641 sample and the Trapezium sample.

As mentioned in \S~\ref{sec:sample}, since we can identify the higher-mass stars through large extinctions and we have a lower limit to the 
number of the low-mass stars, we are overestimating the ratio of high-to-low mass stars and therefore underestimating the significance of 
the result. In addition, up to half of the A stars can actually be background objects. If we account for these A stars, the high-mass bin in 
L1641 would have a smaller number of stars in the $ K < 8$ and $K < 9$ cases. Therefore we expect the P value to be even smaller after 
accounting for A star contamination. 

\section{Discussion}\label{sec:discussion}

\subsection{M$_{ecl}$ - m$_{max}$ relation in Orion A}

\citet{Weidner06} and \citet{Weidner10} compiled a list of Galactic clusters and propose a relation between the most massive star (m$_
{max}$) in a cluster and the cluster mass (M$_{ecl}$). \citet{Bonnell04} also proposed a similar relation based on simulations of 
competitive accretion simulations.In their simulations, the most massive star tends to form in the center of a cluster and gains the majority 
of its mass from the infalling gas onto the cluster, which is accompanied by newly formed low mass stars. Therefore, the formation of the 
high-mass stars correlates with high stellar surface density. Similarly, \citet{Elmegreen04} also suggested that high mass stars can be 
formed from gravitational focused gas accretion in high-density clouds and therefore explain the steep IMF in the field. 
The \citet{Weidner06} and \citet{Bonnell04} M$_{ecl}$ - m$_{max}$ relations 
are almost identical for clusters up to a few thousand \Msun, which suggests that, even though the authors did not explicitly quantify such 
a relation, the upper-mass IMF should also depend on the density. Therefore, qualitatively speaking, the M$_{ecl}$ - m$_{max}$ relation is 
compatible with the environmental density dependence we find in Orion A. 

Quantitatively, whether L1641 follows this M$_{ecl}$ - m$_{max}$ relation is less clear and depends on what we define as a cluster. If we 
consider L1641-s as a whole, its most massive star has a smaller mass than what is expected 
from the M$_{ecl}$ - m$_{max}$ relation. The L1641-s has a total stellar mass of $\sim$1000\Msun, but
 the most massive star is only $\sim $7\Msun. However, if we consider the denser grouping of tens of stars in L1641 individually, it would 
be consistent with the M$_{ecl}$ - m$_{max}$ relation found by \citet{Weidner06}.

\subsection{Challenges in studying the density dependence of the IMF}

Despite the apparent deficiency of high-mass stars in low-density regions, there are few conclusive results that demonstrate the IMF in 
low-density regions is different from that of clusters. Comparing the upper-mass IMF in a low-density star-forming region is difficult for
many reasons, including the small number of high-mass stars intrinsic to the shape of the IMF, the significant line-of-sight contamination 
and the potentially different completeness in regions of different density.

In this work, we compared the largest nearby low-density and clustered star-forming regions and found moderately significant results. On
one hand, this confirms that the IMF is not vastly different in regions of different density; on the other hand, the significance could be 
increased by deeper studies of the extincted population.  We have chosen NOT to use an estimate of the significance that would be 
achieved if we made any correction for incompleteness in the extincted low-mass population to be conservative.   Further study on
the fainter members could make our result even stronger.   In addition, a complete radial velocity survey of the AFG stars would be useful 
to test IMF differences down to lower masses.

\section{Conclusions}\label{sec:conclusion}

We conducted a survey of the intermediate-mass stars in L1641, the lower-density star-forming region south of the ONC, aimed at testing
whether the apparent deficiency of high-mass stars in low-density regions is statistically significant. This study complements the 
low-mass survey presented in Paper I by adding 57 stars in the range of B4 to K4.

In Paper I,  we found that the lack of O and early B stars in L1641 is inconsistent with the \citet{Kroupa01} and \citet{Chabrier05} IMFs. In
this work, our sample of intermediate-mass stars improves our ability to compare directly the ratio of low-mass stars to high-mass stars 
with the ONC.
In particular, we use Fisher's exact test to compare the spectral type distribution of L1641 to that of the ONC from \citet{Hillenbrand97} and 
the K-band luminosity function of L1641 directly to those in the central region of the ONC, the Trapezium \citep{Muench02}. The tests 
indicate a probability of only 3\% that the ONC and the southern region of L1641 (Dec$<$-6.5) were drawn from the same population, 
supporting the hypothesis that the upper mass end of the IMF is dependent on environmental density. 

\acknowledgments
We acknowledge a helpful report from an anonymous referee.
This work was supported in part by NSF grant AST-1008908 and the University of Michigan.
We made use of SAO/NASA Astrophysics Data System. This paper uses data obtained at the MMT Observatory, a joint facility of the Smithsonian 
Institution and the University of Arizona and the 6.5 meter Magellan Telescopes located at Las Campanas Observatory, Chile. 

{\it Facilities:} \facility{MMT (Hectospec)}, \facility{Magellan:Baade (IMACS)}, \facility{Hiltner (OSMOS)}, \facility{Spitzer (IRAC)},
\facility{FLWO:2MASS}

\bibliography{./Bib}

\end{document}